\documentclass[preprint]{aastex}

\newcommand{\eg}{{\it e.g.}}
\newcommand{\etal}{{\it et al.}}

\newcommand{\aasa}{American Astronomical Society Meeting Abstracts}

\newcommand{\ApSS}{Astrophys. and Space Sci.}
\newcommand{\ieeep}{IEEE Proceedings}
\newcommand{\ieeetap}{IEEE Transactions on Antennas and Propagation}
\newcommand{\jai}{Journal of Astronomical Instrumentation}
\newcommand{\lasp}{Leaflet of the Astronomical Society of the Pacific}
\newcommand{\obs}{The Observatory}

\begin{document} 


\shorttitle{Antenna System for Long Wavelength Radio Astronomy}
\shortauthors{Hicks, \etal}

\title{A wide-band, active antenna system for long wavelength radio astronomy} 

\author{Brian C.~Hicks}
\affil{Naval Research Laboratory, Code 7213, Washington, DC 20375-5320}
\email{Brian.Hicks@nrl.navy.mil}

\author{Nagini Paravastu-Dalal\altaffilmark{1}}
\affil{Northrop Grumman Aerospace Systems, 1 Space Park, Mailstop R4/1400AH, Redondo Beach, CA 90278}
\email{Nagini.Dalal@ngc.com}

\author{Kenneth P.~Stewart}
\affil{Naval Research Laboratory, Code 7211, Washington, DC 20375-5320}
\email{Kenneth.Stewart@nrl.navy.mil}

\author{William C.~Erickson}
\affil{University of Tasmania, Churchill Ave., Sandy Bay, Tasmania 7005, Australia}
\email{bill.erickson@utas.edu}

\author{Paul S.~Ray}
\affil{Naval Research Laboratory, Code 7655, Washington, DC 20375-5352}
\email{Paul.Ray@nrl.navy.mil}

\author{Namir E.~Kassim}
\affil{Naval Research Laboratory, Code 7213, Washington, DC 20375-5320}
\affil{Namir.Kassim@nrl.navy.mil}

\author{Steve Burns}
\affil{Burns Industries, 141 Canal Street Nashua,  NH 03064}
\email{info@burnsindustriesinc.com}

\author{Tracy Clarke}
\affil{Naval Research Laboratory, Code 7213, Washington, DC 20375-5320}
\email{Tracy.Clarke.ca@nrl.navy.mil}

\author{Henrique Schmitt}
\affil{Naval Research Laboratory, Code 7215, Washington, DC 20375-5320}
\email{Henrique.Schmitt.ctr.br@nrl.navy.mil}

\author{Joe Craig}
\affil{University of New Mexico, 1919 Lomas Blvd. NE, Albuquerque, NM 87131}
\email{joecraig@unm.edu}

\author{Jake Hartman}
\affil{Jet Propulsion Laboratory, MS 138.308, Pasadena, CA 91109}
\email{Jacob.M.Hartman@jpl.nasa.gov}

\author{Kurt W.~Weiler}
\affil{Computational Physics, Inc., 8001 Braddock Rd., Springfield, VA 22151}
\email{Kurt.Weiler@weilerhome.org}

\altaffiltext{1}{Formerly NRL ASEE Postdoctoral Fellow}

\begin{abstract}
We describe an  ``active'' antenna system for HF/VHF (long wavelength) radio astronomy that has been successfully deployed 256-fold as the first station (LWA1) of the planned Long Wavelength Array. The antenna system, consisting of crossed dipoles, an active balun/preamp, a support structure, and a ground screen has been shown to successfully operate over at least the band from 20 MHz (15 m wavelength) to 80 MHz (3.75 m wavelength) with a noise figure that is at least 6 dB better than the Galactic background emission noise temperature over that band. Thus, the goal to design and construct a compact, inexpensive, rugged, and easily assembled antenna system that can be deployed many-fold to form numerous large individual  ``stations'' for the purpose of building a large, long wavelength synthesis array telescope for radio astronomical and ionospheric observations was met.
\end{abstract}


\section{Introduction}

\subsection{Background\label{background}}

Radio astronomy began in 1932 with the discovery of radio emission from the Galactic Center at the relatively long wavelength of 15 m (20 MHz) by Karl Jansky \citep{Jansky32,Jansky33a,Jansky33b,Jansky33c}. This pioneering work was followed by the innovative research of Grote Reber at frequencies ranging from 10 -- 160 MHz (~30 -- 2 m wavelength) in the 1940s that closely tied radio astronomy to the broader field of astronomy and astrophysics \citep{Reber40,Reber44,Reber47,Reber49,Reber50}. 

However, the requirement for impractically large single radio antennas or dishes to obtain resolution at long wavelengths (resolution $\theta \sim \lambda/\rm D$, where $\theta$ is the angular resolution in radians, $\lambda$ is the observing wavelength in meters, and D is the diameter of the observing instrument in meters) quickly pushed the new field of radio astronomy to higher frequencies (shorter wavelengths). In other words, since increasing D was severely limited by cost and mechanical considerations, the only way to achieve better resolution seemed to be to decrease $\lambda$. 

As early as 1946, Ryle and Vonberg \citep{Ryle46,Ryle48} and Pawsey and collaborators \citep{Pawsey46a,Pawsey46b} began to use interferometric techniques consisting of large arrays of simple dipoles or widely separated individual, small dishes to increase the effective  ``D'' without greatly increasing the cost. Even then, distortions introduced into the incoming radio signals by the Earth's ionosphere made imaging at long wavelengths difficult  and appeared to place a rather short upper size limit to ``D'' at frequencies $<100$ MHz (wavelengths $>3$ m) of $\sim 5$ km. Thus, the move to higher frequencies, even for interferometry, continued until, by the 1970s, relatively few long wavelength radio astronomy telescopes were still operating at frequencies $<100$ MHz. Exceptions include the Ukrainian UTR-2 \citep{Braude78}, the 38 MHz survey \citep{Rees90} with the Cambridge Low-Frequency Synthesis Telescope \citep[CLFST; see ][]{Baldwin85,Hales88} and the Gauribidanur Radio Observatory (GEETEE) in India \citep{Shankar90,Dwarakanath95}. 

Another important long wavelength array in the 1970s and 1980s was the ``Tee Pee Tee'' (TPT) Clark Lake array built by William C. (Bill) Erickson on a dry lake in the Anza-Borrego desert east of San Diego, CA \citep{Erickson82}. The TPT was also limited to a maximum baseline ``D'' of ~3 km because of concerns about ionospheric distortion.  For a more complete history of the Clark Lake Radio Observatory (CLRO) TPT array, its precedents and follow-on arrays,  see  \cite{Kassim05a}. Due to lack of funding, the CLRO TPT was decommissioned in the late 1980s and dismantled in the early 1990s.

\subsection{Modern long wavelength arrays\label{modernarrays}}

A significant development in radio astronomy in the early 1980s known as Self Calibration or ``Self-Cal''  \citep{Pearson84} finally showed that the ``ionospheric barrier'' could be overcome. Self-Cal makes use of one or more sources in the field of view to monitor instrumental, atmospheric, and ionospheric changes on short time scales so that they can be removed during the data reduction processes. This new data handling method allowed astronomers to contemplate high resolution imaging through the ionosphere at frequencies $<100$ MHz and revitalized the field of long wavelength radio astronomy.

The defining paper for the start of this ``modern era'' of long wavelength arrays is probably the one by \cite{Perley84} that made a strong scientific and technical case for a long wavelength synthesis array operating at $\sim 75$ MHz located near the Very Large Array (VLA) of the National Radio Astronomy Observatory (NRAO\footnote{The National Radio Astronomy Observatory is a facility of the National Science Foundation operated under cooperative agreement by Associated Universities, Inc.}) and making use of the VLA's existing infrastructure. Although this proposal was not funded or constructed, it certainly influenced subsequent plans for long wavelength synthesis arrays for radio astronomy.

In particular, Self-Cal, the CLRO TPT array, and the \cite{Perley84} concept led directly to the proposal and installation, with Naval Research Laboratory (NRL) and NRAO financial and technical support, of a 74 MHz  (4 m) feed and receiver system on one of the VLA 25m dishes in 1991, eight VLA dishes by 1994, and all 27 VLA dishes by 1998. When fully implemented, this ``4-band'' system became a universally available user band \citep[see, \eg,][]{Kassim93,Kassim05a,Kassim07}. 

The international success of this 4-band system demonstrated both the scientific richness of the $<100$ MHz frequencies and the possibilities for using these new data reduction techniques for overcoming the previous limitations of relatively short ($<5$ km) baseline lengths to obtain arc-second resolution from ground-based, long wavelength imaging arrays.

Such resounding success also led to the concept of a dedicated low frequency ($\nu < 100$ MHz) Long Wavelength Array (LWA) in the late 1990s \citep{Kassim98} and, shortly thereafter, NRL, MIT, and ASTRON (Netherlands) formed the LOFAR Consortium to further develop the plans that are now present in the Dutch-led Low Frequency Array \citep[LOFAR;][]{Rottgering03,Wijnholds11}, the somewhat higher frequency MIT-led Murchison Widefield Array \citep[MWA;][]{Lidz08}, and the University of New Mexico/NRL-led Long Wavelength Array \citep[LWA;][]{Kassim98,Kassim05a,Kassim05b,Ellingson09a}. For technical reasons, such long wavelength arrays need a large number ($\ge10,000$) of electromagnetic-wave receptors \citep[see, \eg,][]{Ellingson05,Ellingson09a,Ellingson11}  so that the development of a cheap, easily deployable antenna system that is Galactic background noise limited\footnote{At long wavelengths (particularly at frequencies $<100$ MHz), the Galactic background radio emission is the ultimate limit on the effective noise temperature of any radio receiving system. Thus, the active balun/preamp (frontend) should provide enough gain that any noise contributed by components following it is negligible. Also, in order to have the frontend noise itself not raise the total system noise temperature much above that of the fundamental Galactic background limit, it must have a noise temperature significantly below that of the Galactic background at the observing wavelengths of interest. While a cooled, very low noise temperature receiver might be desirable, it is clearly not practical to build  such a unit at reasonable cost when it must be reproduced thousands of times in a large, long wavelength array. As a compromise between a ``perfect'' frontend and an affordable frontend, the specification given in Table \ref{tbl-LWA-spec} for Sky Noise Dominance (SND) was chosen such that the frontend should have a noise temperature better than 6 dB below the Galactic background noise temperature over the principal band of interest from 20 -- 80 MHz. At 6 dB below the Galactic background, the increased integration time to reach a given sensitivity is only $\sim57$\% more than with a perfect, noiseless balun/preamp \citep{Erickson05}. This was considered to be acceptable.} (see Section \ref{SystemPerform}) is vital. 

Initial prototyping of such an antenna system was carried out while NRL was still part of the LOFAR Consortium and tested on the NRL LOFAR Test Array \citep[NLTA; ][]{Stewart04,Stewart05}. The NLTA, an 8-element, long wavelength array employing active, droopy-dipole, fat antennas and operating between $\sim 15$ MHz ($\sim20$ m) to $\sim 115$ MHz ($\sim2.6$ m), provided valuable experience leading to next stage of prototyping with the Long Wavelength Demonstrator Array \citep[LWDA; ][]{Lazio10}. The LWDA was  a 16-element long wavelength synthesis array operating between 60 MHz (5 m wavelength) and 80 MHz (3.75 m wavelength) constructed by the Applied Research Laboratories of the University of Texas, Austin in collaboration with, and funding from, NRL. Lessons from these two prototyping efforts led to the improved design for the active antenna system for the LWA that is the focus of this paper.

\subsection{The Long Wavelength Array (LWA)\label{LWA}}

The LWA is a long wavelength radio astronomy synthesis array now under construction. It is designed to enable new research in the largely unexplored frequency range of $\sim 20$ -- 80 MHz (15 m -- 3.75 m wavelength) with reduced sensitivity both above and below that range. When completed, the full LWA will require $>10,000$ full polarization, crossed dipole antenna elements organized into $\sim50$ ``stations,'' each station consisting of 256 antenna elements distributed over an ellipse $\sim100$ m E/W by $\sim110$ m N/S with a quasi-random placement. The entire $\sim50$ station synthesis array will eventually be spread over an area roughly 400 km in diameter centered in the state of New Mexico. A possible antenna concept was given already by \cite{Hicks02} and a compact description of a concept array can be found in \cite{Kassim05b}. Specifications from this latter paper are included here in Table \ref{tbl-LWA-spec} for easy reference. \cite{Clarke09} also provides an excellent discussion of requirements and specifications. For an updated description of the actual parameters of the LWA1, see \cite{Ellingson12} and \cite{Taylor12}. 

\begin{figure}
\plotone{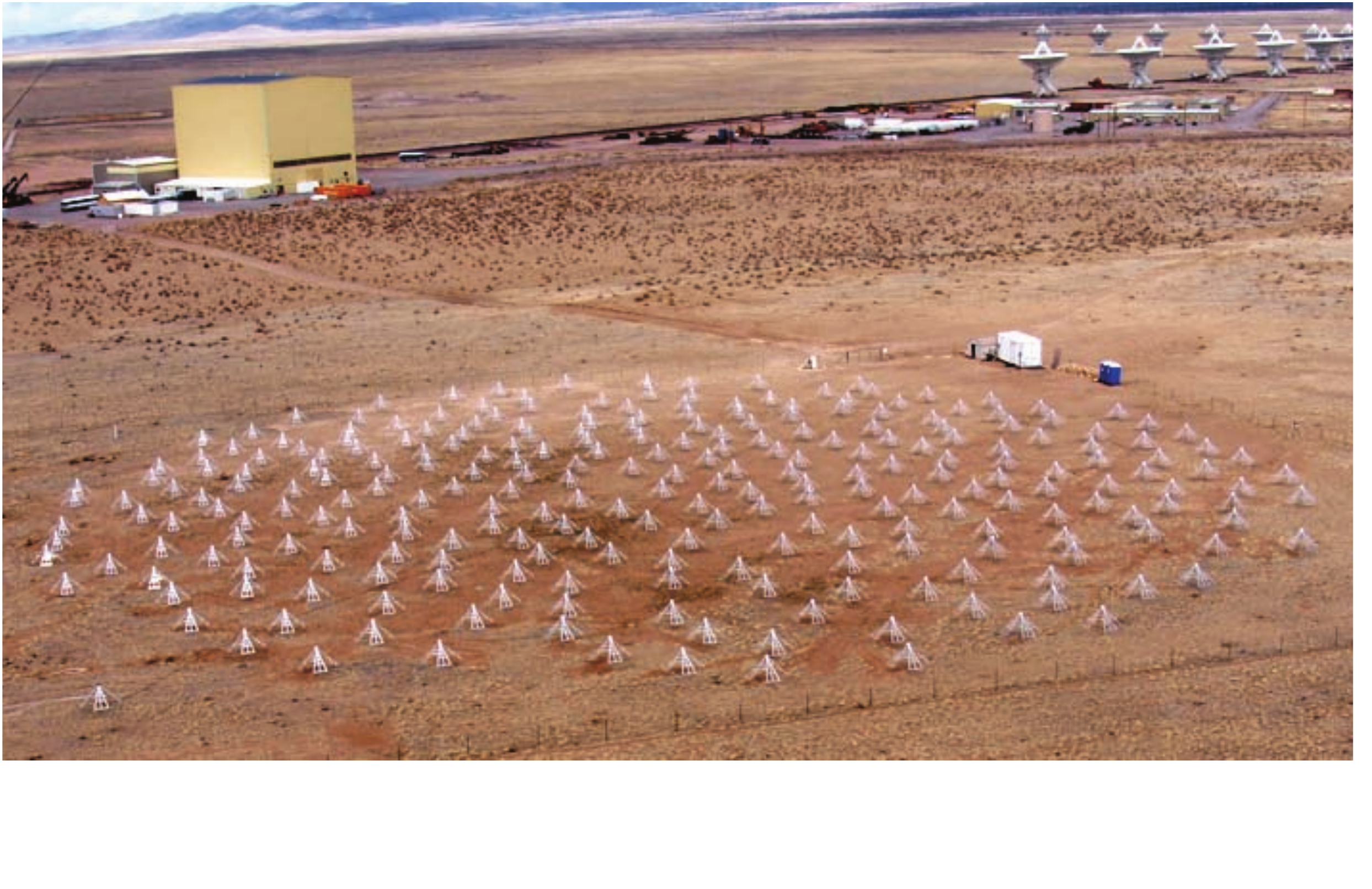}
\caption{An aerial photograph of the first station for the Long Wavelength Array (LWA1) located near the center of the National Radio Astronomy Observatory's (NRAO's) Very Large Array (VLA) about 100 km west of Socorro, NM. The array consists of 256 active antenna stands (described in this paper), located in a quasi-random distribution within an ellipse of dimensions $\sim100$ m E/W by $\sim110$ m N/S. The station (designated for its stand alone use as the LWA1 Radio Observatory) is currently performing observations resulting from its first call for proposals in addition to a continuing program of commissioning and characterization observations \citep[see ][]{Ellingson12,Taylor12}. Note that a number of 25 m dishes of the VLA are visible in the upper right hand corner.}
\label{arraypix}
\end{figure}

To meet these stringent, and often conflicting requirements at reasonable cost, and drawing on our prototyping experience described above, we chose an electrically short, relatively  fat, droopy-dipole design similar to that shown to be effective on our NLTA and LWDA prototypes with an amplified or ``active'' balun/pre-amplifier at the apex of the dipole arms. Of course, a support for these droopy dipoles that was simple and easy to install had to be designed and, to stabilize the properties of the ground under the antenna against changes in, \eg, moisture content (such as rain), an inexpensive and rugged ``ground screen'' had to be included. All of these elements of dipole antenna (ANT), front-end electronics (FEE), support stand (STD), and ground screen (GND) work as a coupled system and had to be designed together. In this paper we describe the electrical and mechanical properties of these four components that are already designed, built, and deployed 256-fold as the first LWA1 station (designated for its stand alone use as the LWA1 Radio Observatory). The LWA1 Radio Observatory is currently performing observations resulting from its first call for proposals in addition to carrying out a continuing program of commissioning and characterization observations \citep[see Figure \ref{arraypix} and][]{Ellingson12,Taylor12}. 

It should be noted that the ANT/FEE/STD/GND system described here has also proven to be versatile enough to draw interest for other radio astronomy applications for both national observatory  [\eg, Nan\c{c}ay Observatory in France \citep{Girard11}] and university groups [\eg. the Low Frequency All Sky Monitor, LOFASM, of the University of Texas, Brownsville \cite[][]{Miller12,Rivera12} and the Universidad Nacional Autonoma de Mexico, UNAM].

\section{An antenna system for HF/VHF radio astronomy\label{ANT-FEE-STD-GND}}

For simplicity, we break the discussion of this ``antenna system'' into four parts: the  crossed-dipole antenna (ANT), the front-end electronics (FEE), the support stand (STD), and the ground screen (GND). We will first describe each of these separately, and then report on the electrical response of the entire system.

\subsection {Antenna - ANT\label{ANT}}

\subsubsection{Geometry\label{antgeom}}

Extensive simulations, measurements, and prototyping were carried out to determine a satisfactory antenna design considering cost, weight, mechanical stability, wind resistance, ease of fabrication, and RF performance. These were briefly mentioned above and are discussed further in  \cite{Paravastu07a,Paravastu08a,Paravastu08c}. 

From electrical considerations, it was clear that the antenna elements had to be broad in shape to improve the inherent bandwidth characteristics over those of a simple, thin-wire dipole.  Furthermore, the elements had to slope downward at $45\degr$ to improve the sky coverage over that of a simple, straight dipole. Drawing on the NLTA and LWDA experience, initial tests were with broad, flat sheets of aluminum, roughly 1.75 m long $\times$ 0.42 m wide, sharply tapered to a feed point.  While the performance of these ``Big Blades'' was encouraging \cite[see, \eg,][]{Erickson06,Kerkhoff07a,Paravastu08b}, it was estimated that the cumbersome size, high wind resistance, and large metal content/cost were unsatisfactory \citep{Paravastu08c}. 

\begin{figure}
\plotone{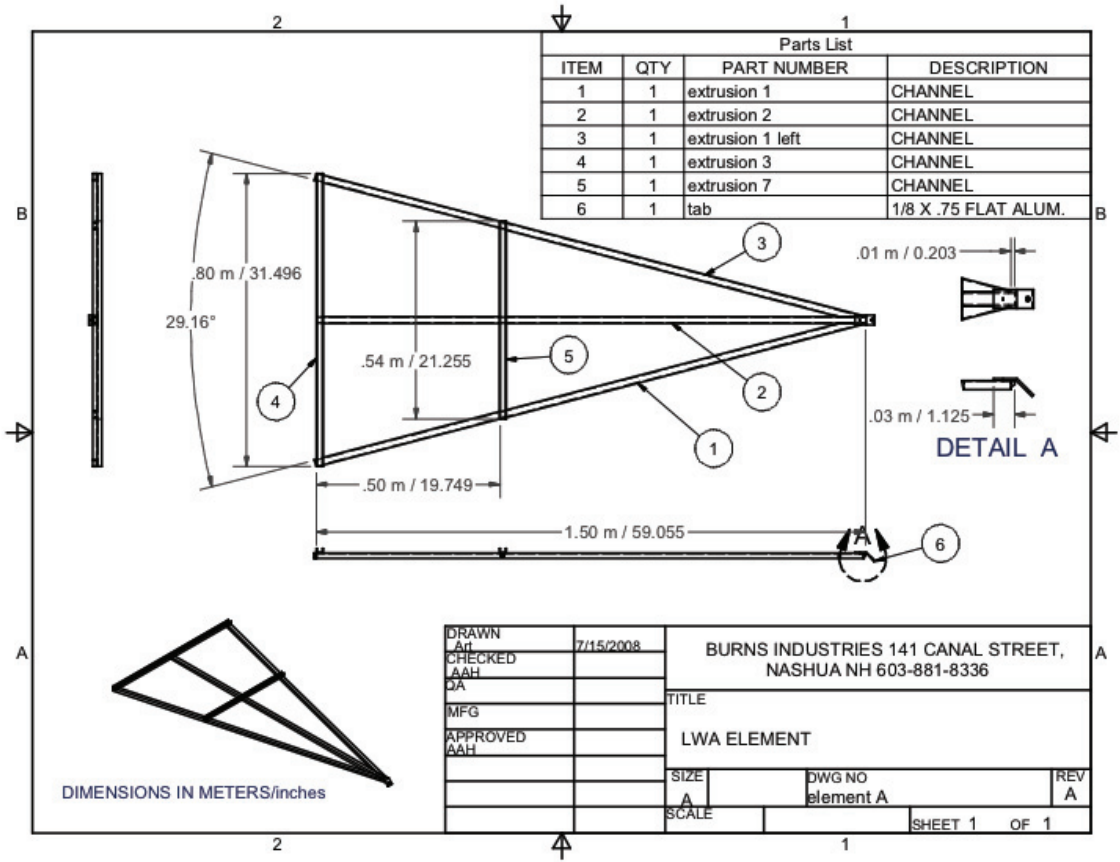}
\caption{Mechanical design of one arm of the tied fork with crosspiece dipole.\label{ANT-Figure2-2}}
\end{figure}

Next a series of ``frame'' antennas using aluminum angle pieces, with and without vertical and horizontal cross pieces, and with and without mesh covering, were considered. Again, the electrical results were satisfactory, but the metal cost remained high. Finally, just a triangular frame of aluminum angle pieces with a single vertical bar (known as the ``tied fork'') and a single horizontal crosspiece for increased stiffness, was chosen for the final design. This selection process is described in  \cite{Paravastu07a,Paravastu08c} and one arm of this  ``tied fork with crosspiece'' is shown in Figure  \ref{ANT-Figure2-2}. The vertical height of this triangle is 1.50 m and the base of the triangle is 0.8 m. The distance between the feed points on the FEE (see Section \ref{FEE}) unit is 9.0 cm and the apexes of the triangular ANT elements are separated by about 13.2 cm. Numerous simulations \citep[][using the the experimental software package Numerical Electromagnetics Code NEC\footnote{www.nec2.org}-4.1 provided by the Lawrence Livermore National Laboratory]{Paravastu07a} were carried out and field tests were performed  \citep{Paravastu08c} on both early and later designs. The simulation and field test results indicated that the ``tied fork with crosspiece'' yielded the best compromise for low cost, high mechanical stability, and good electrical performance. 

\subsubsection{Electrical performance\label{ANTperform}}

\begin{figure}
\plottwo{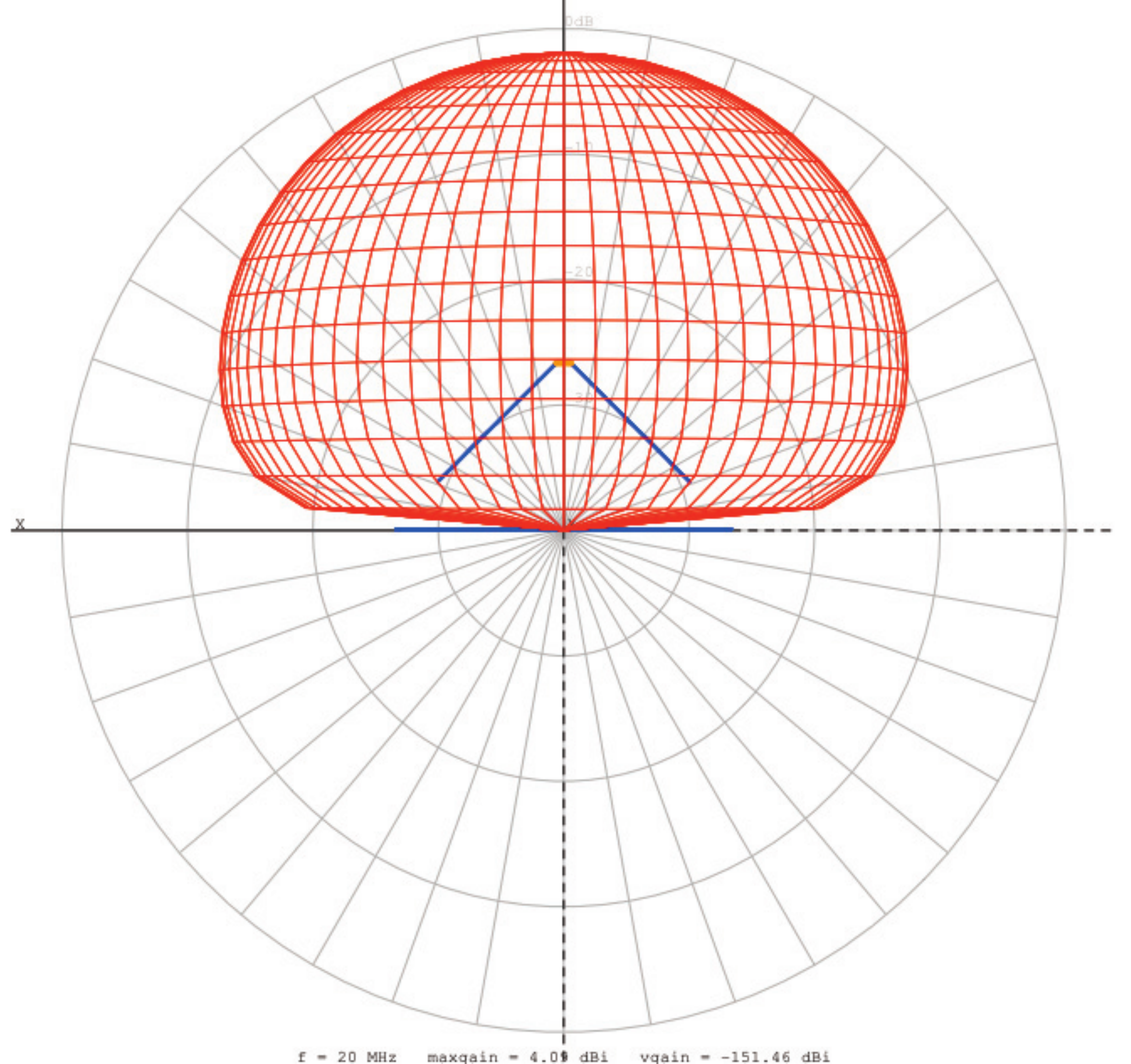}{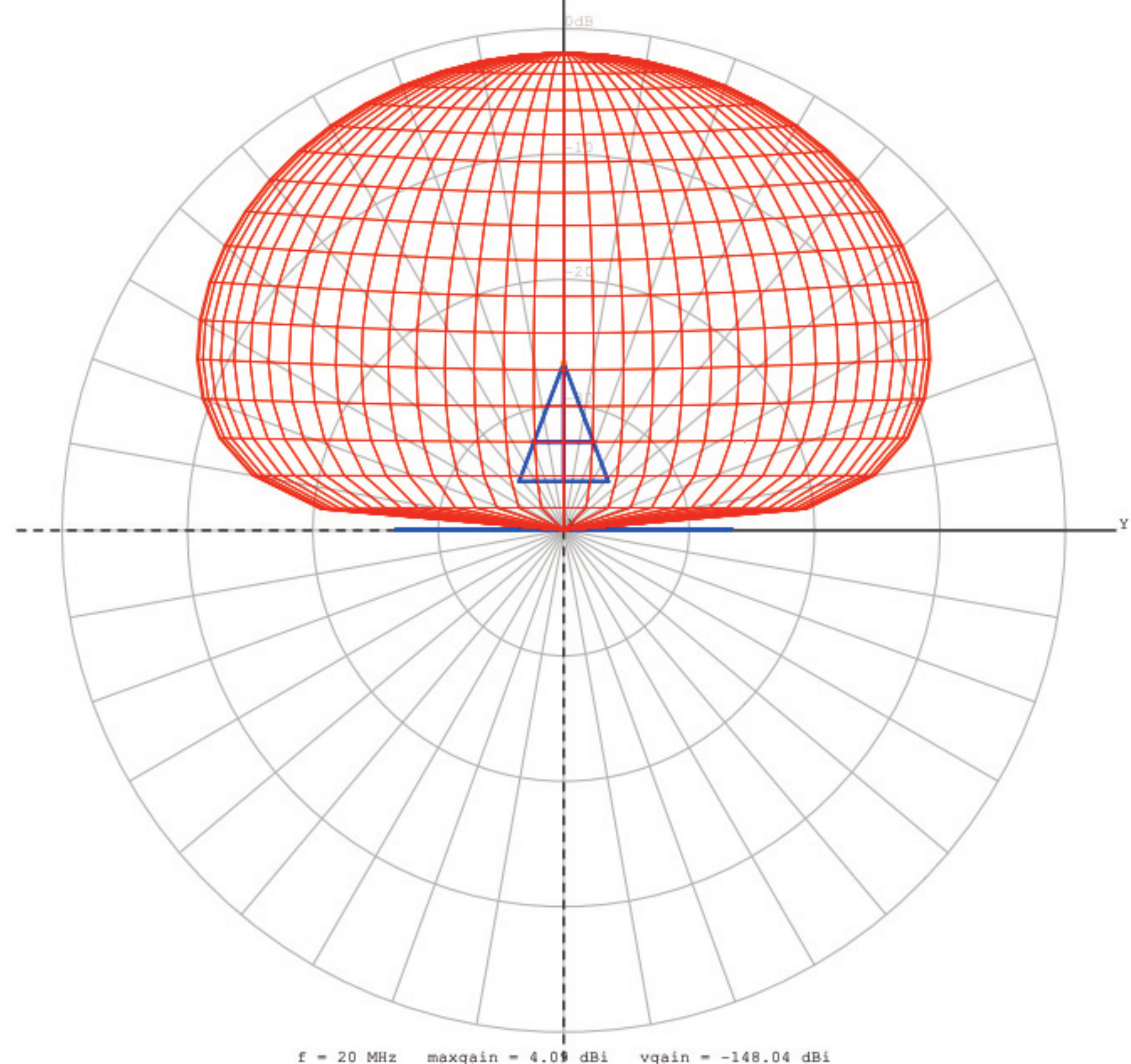}
\plottwo{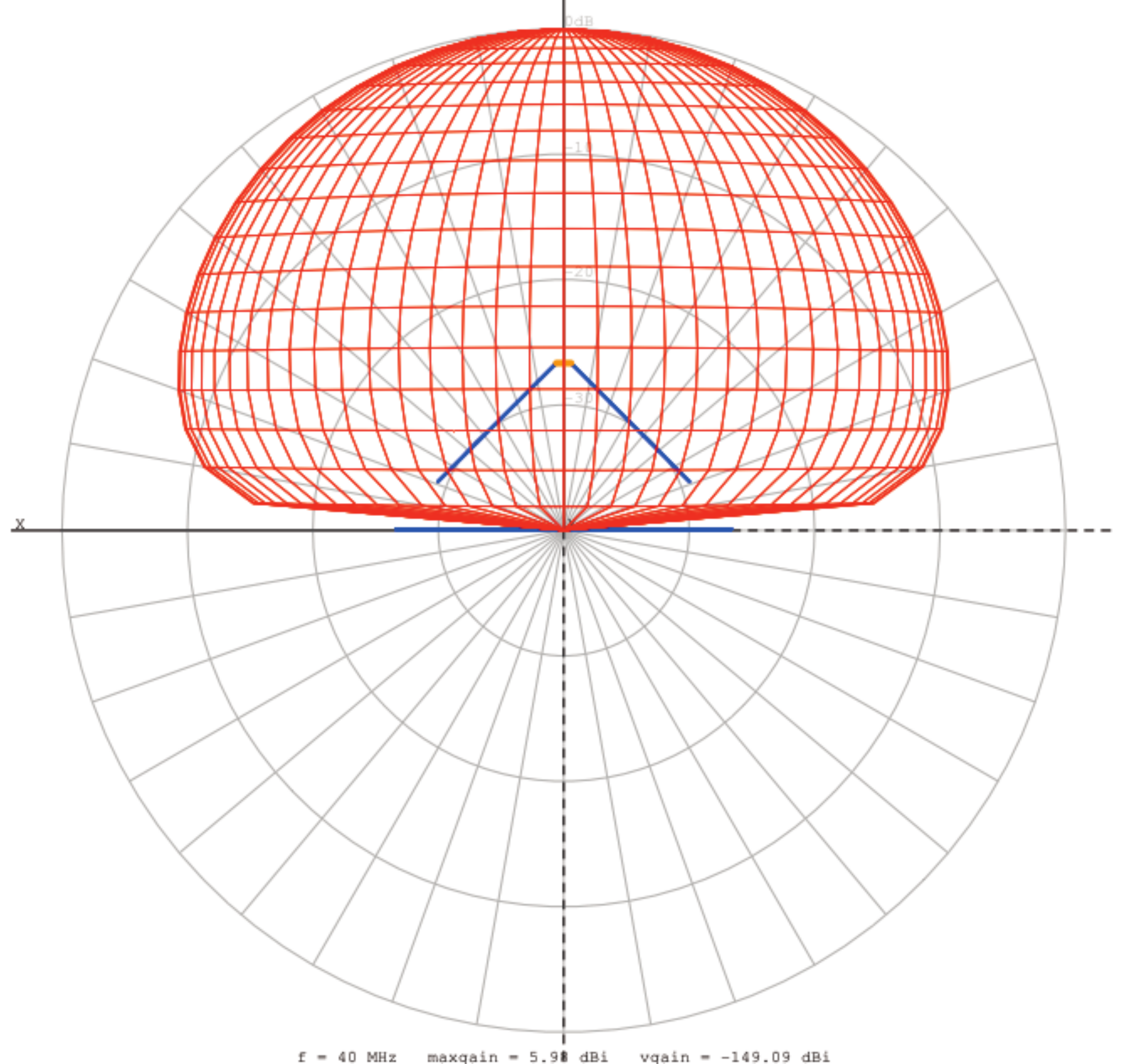}{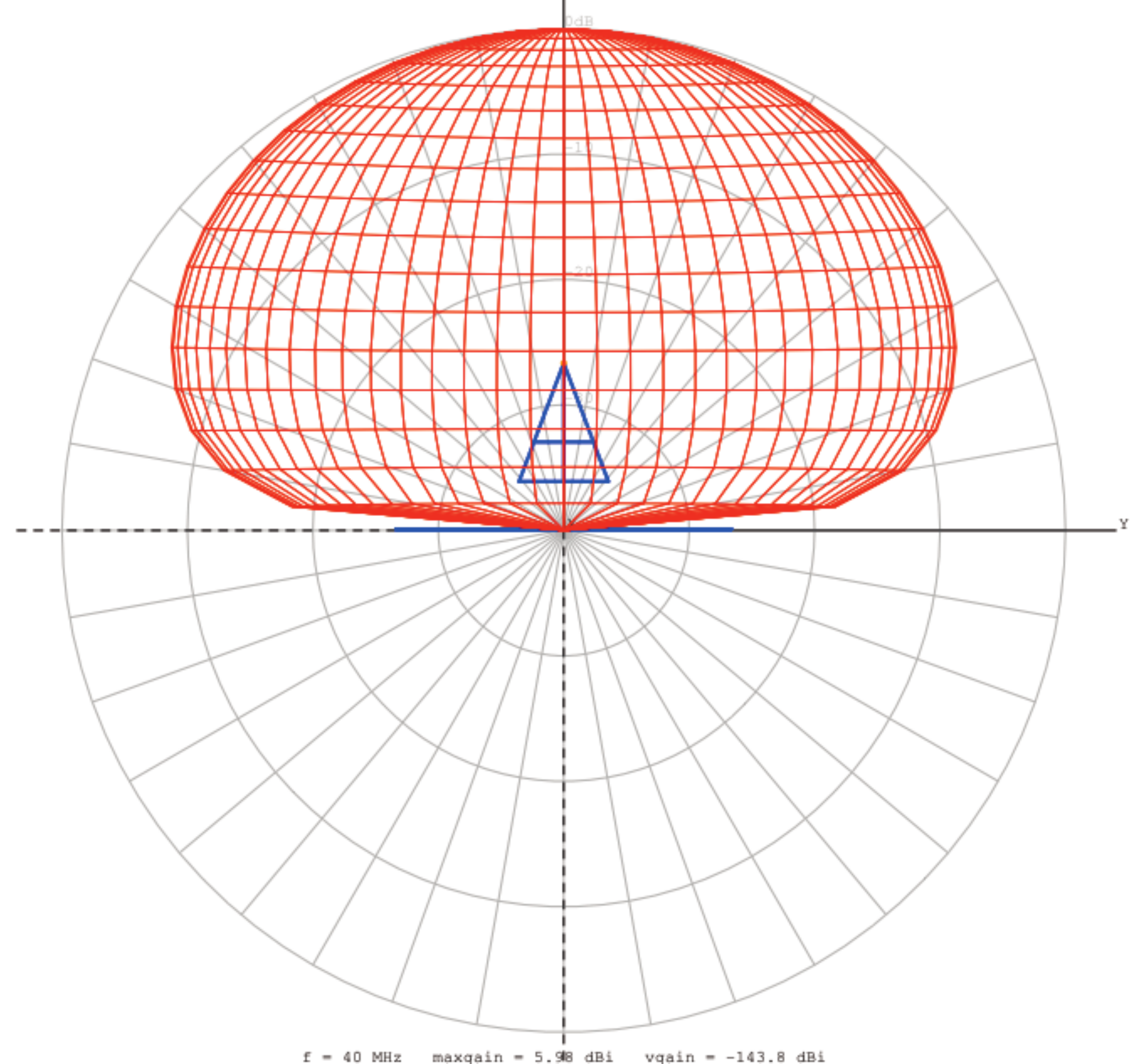}
\caption{Simulated E and H plane power patterns at 20 MHz (top) and 40 MHz (bottom) for the ``tied fork with crosspiece'' antenna. The scale is logarithmic total power with a normalization of unity at the zenith and $-$10 dB per radial division below that. E-plane patterns are on the left and H-plane patterns are on the right. The blue vertex in the left hand figures and the blue crossed/barred triangles in the right hand figures represent the antenna viewed edge on and front on, respectively. The blue line along the horizontal axis in all figures represents the ground screen viewed edge on. It should be noted that even though the simulations were initially carried out with a single polarization to enhance computing speed, a final check was always made with both polarizations in place to insure that the presence of the other polarization did not change the results.}
\label{ANT-Figure2-4}
\end{figure}

\begin{figure}
\plottwo{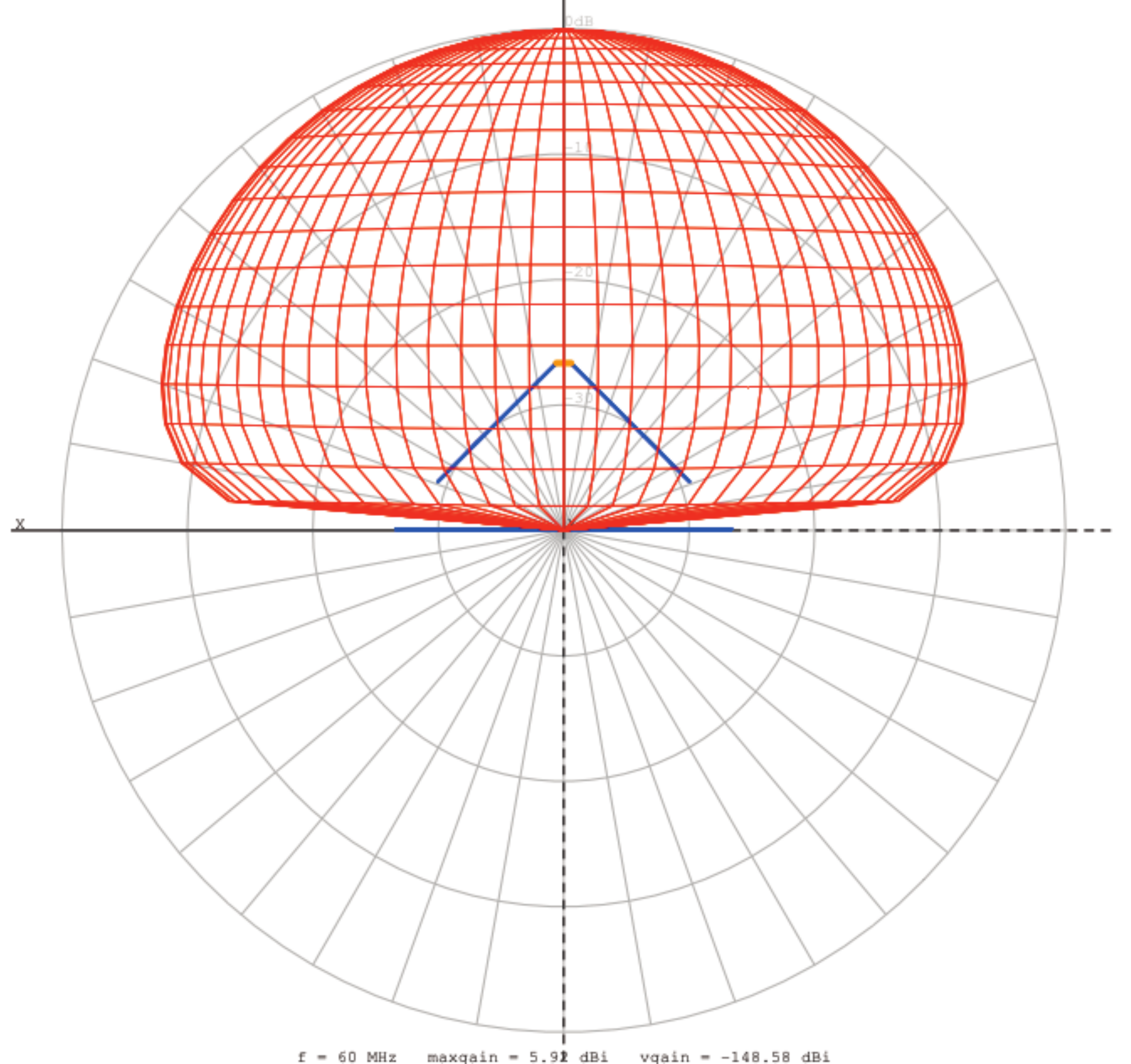}{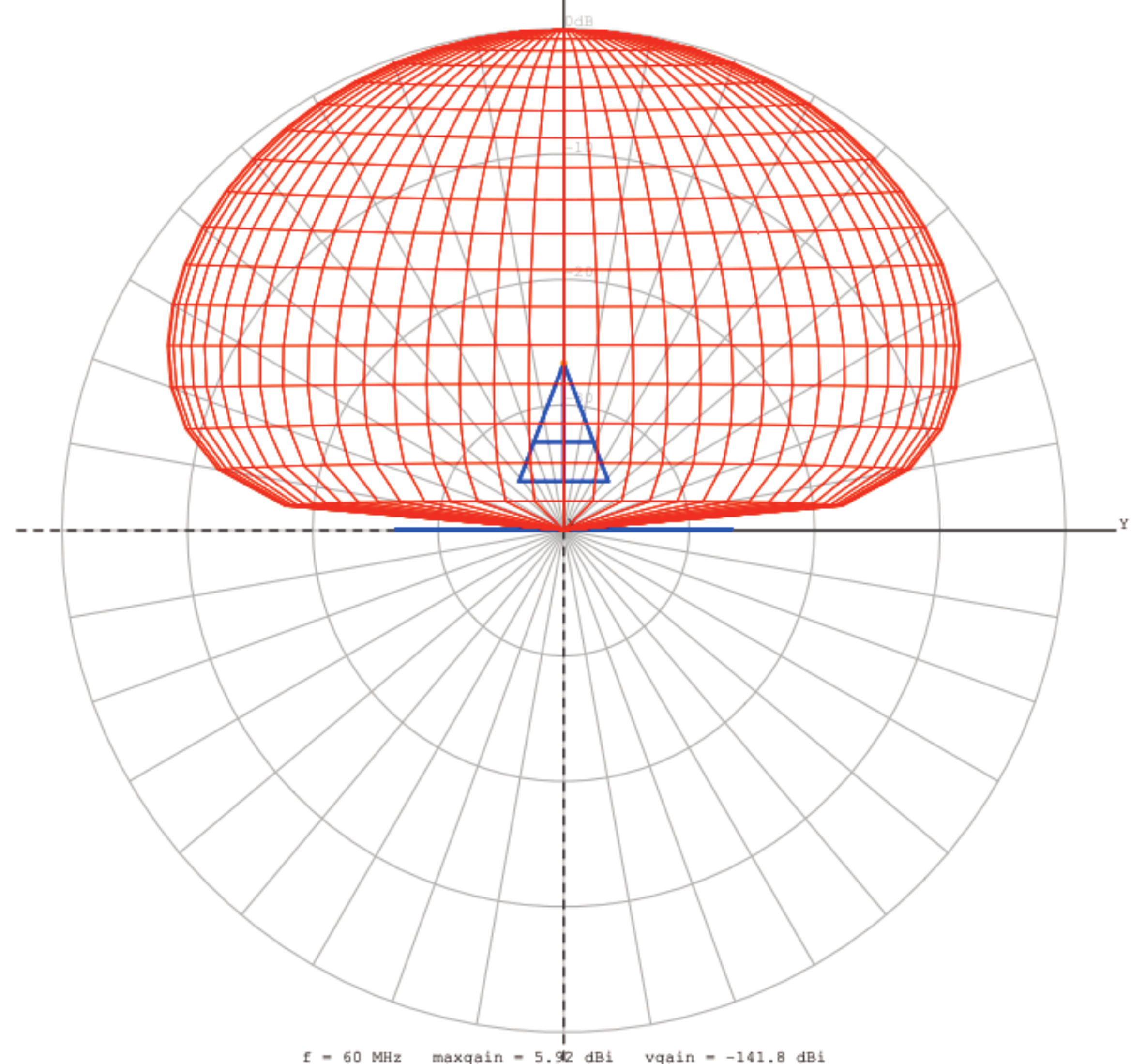}
\plottwo{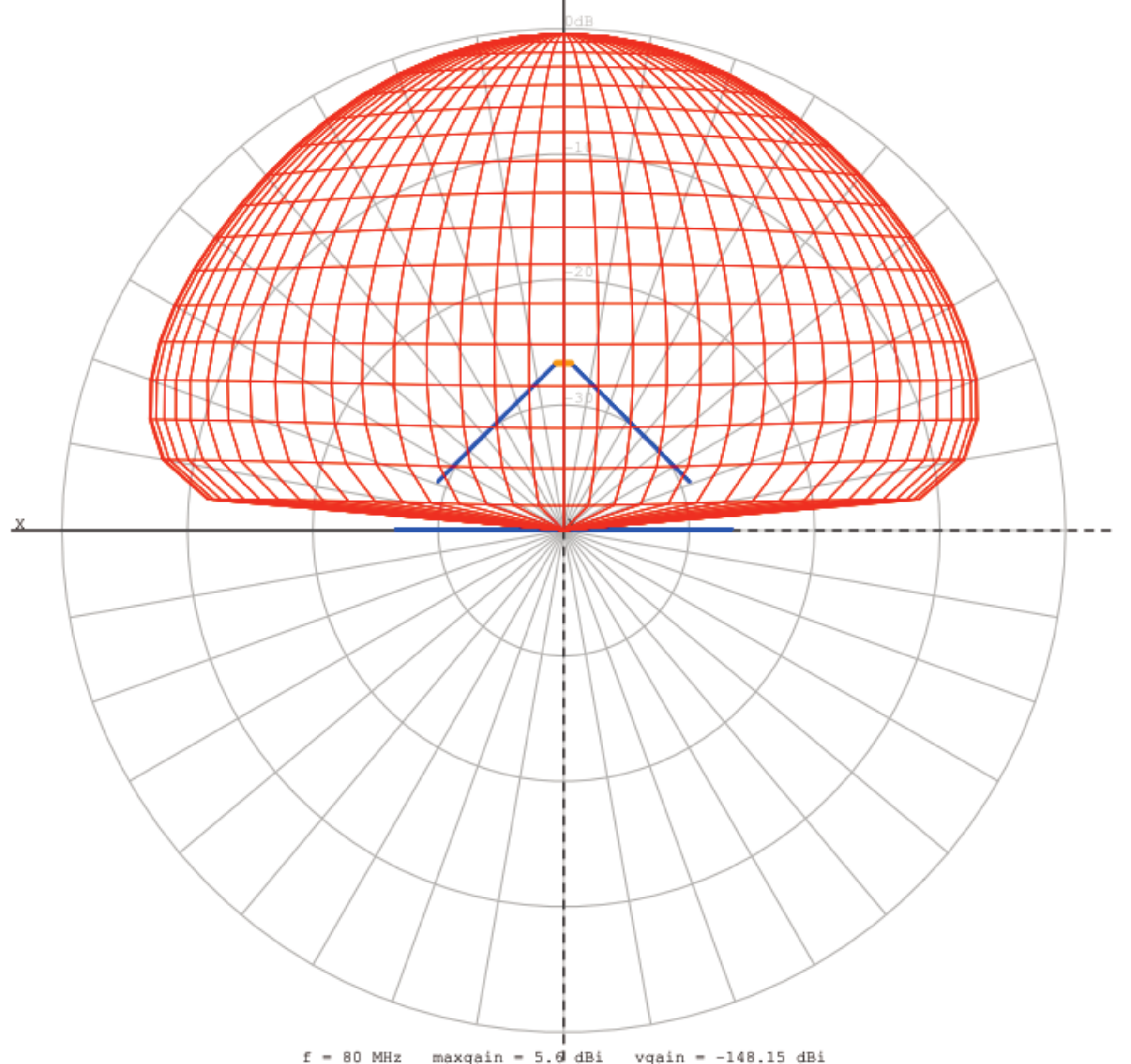}{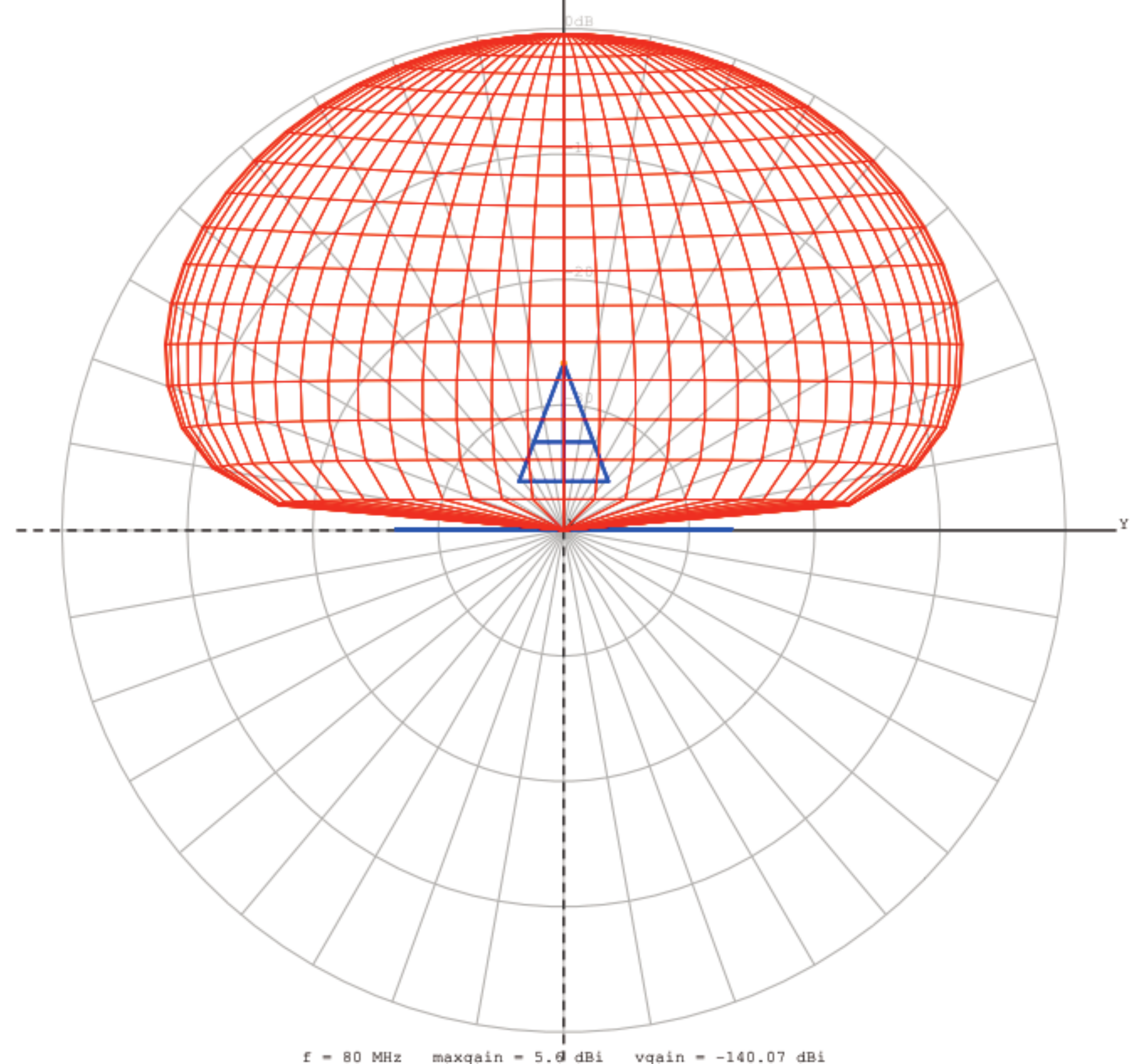}
\caption{As for Figure \ref{ANT-Figure2-4}, simulated E and H plane power patterns at 60 MHz (top) and 80 MHz (bottom). The scale is logarithmic total power with a normalization of unity at the zenith and $-10$ dB per radial division below that.}
\label{ANT-Figure2-5}
\end{figure}

\begin{figure}
\plottwo{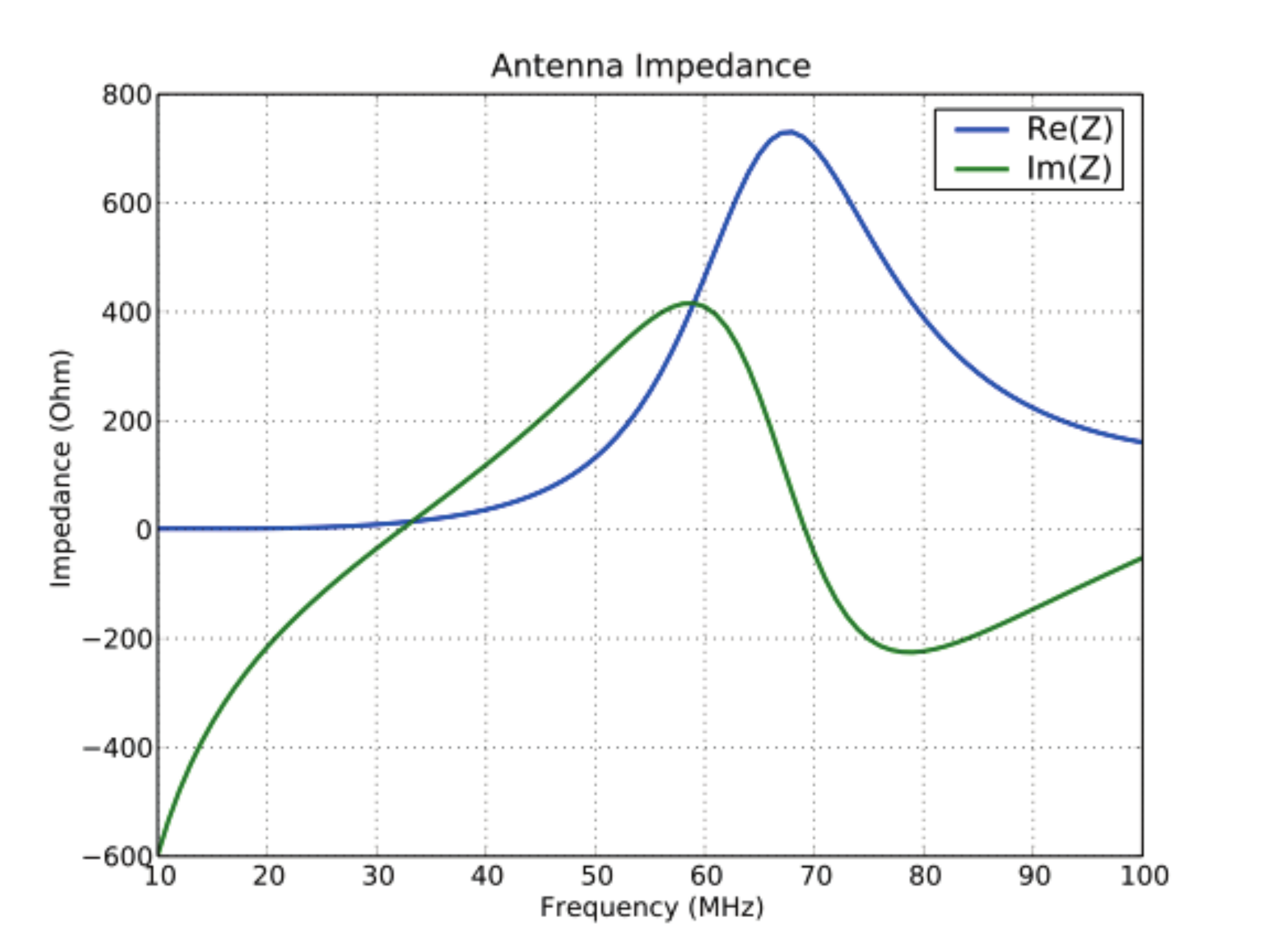}{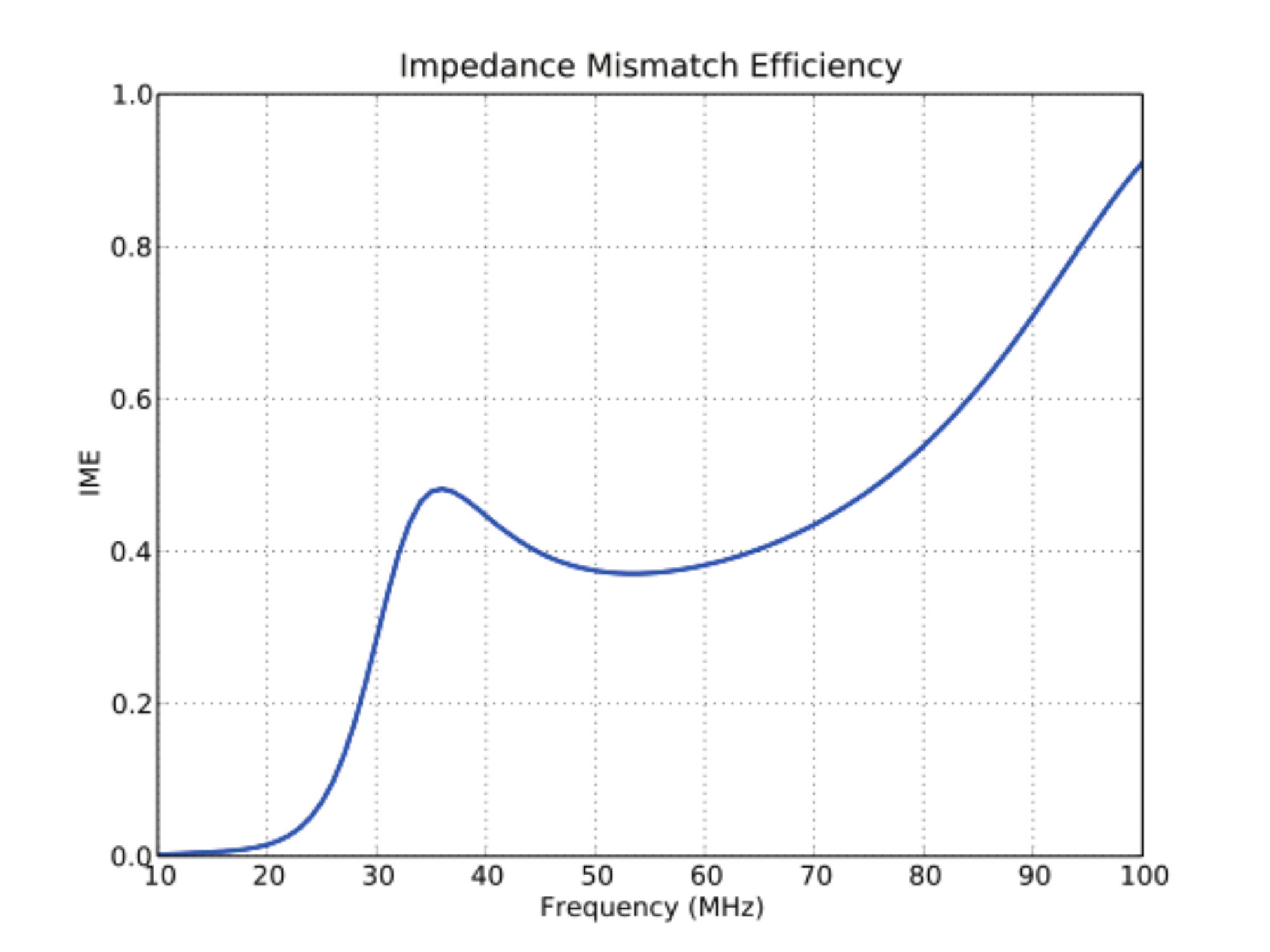}
\caption{(left) Antenna terminal impedance (Z) and (right) impedance mismatch efficiency (IME; 1-abs[(Z-Z$_0$)/(Z+Z$_0$)]) where Z is the impedance of the antenna and Z$_0$ is the impedance of the preamp.}
\label{ANT-Figure2-6}
\end{figure}

The simulated E- and H-plane patterns over a range of frequencies are shown in Figures \ref{ANT-Figure2-4} and \ref{ANT-Figure2-5} and summarized in Table \ref{tbl-ANTpatt}. Actual measurements discussed in \cite{Hartman09a} indicate that the simulated and measured beam patterns correspond to better than 1 dB over almost all of the sky. The predicted impedance characteristics are shown in Figure \ref{ANT-Figure2-6} with the antenna terminal impedance (Z) shown on the left and the impedance mismatch efficiency (IME; the fraction of the power at the antenna feed point that is transferred to the preamp) shown on the right. 

\subsection{Front-end electronics - FEE\label{FEE}}

\begin{figure}
\plotone{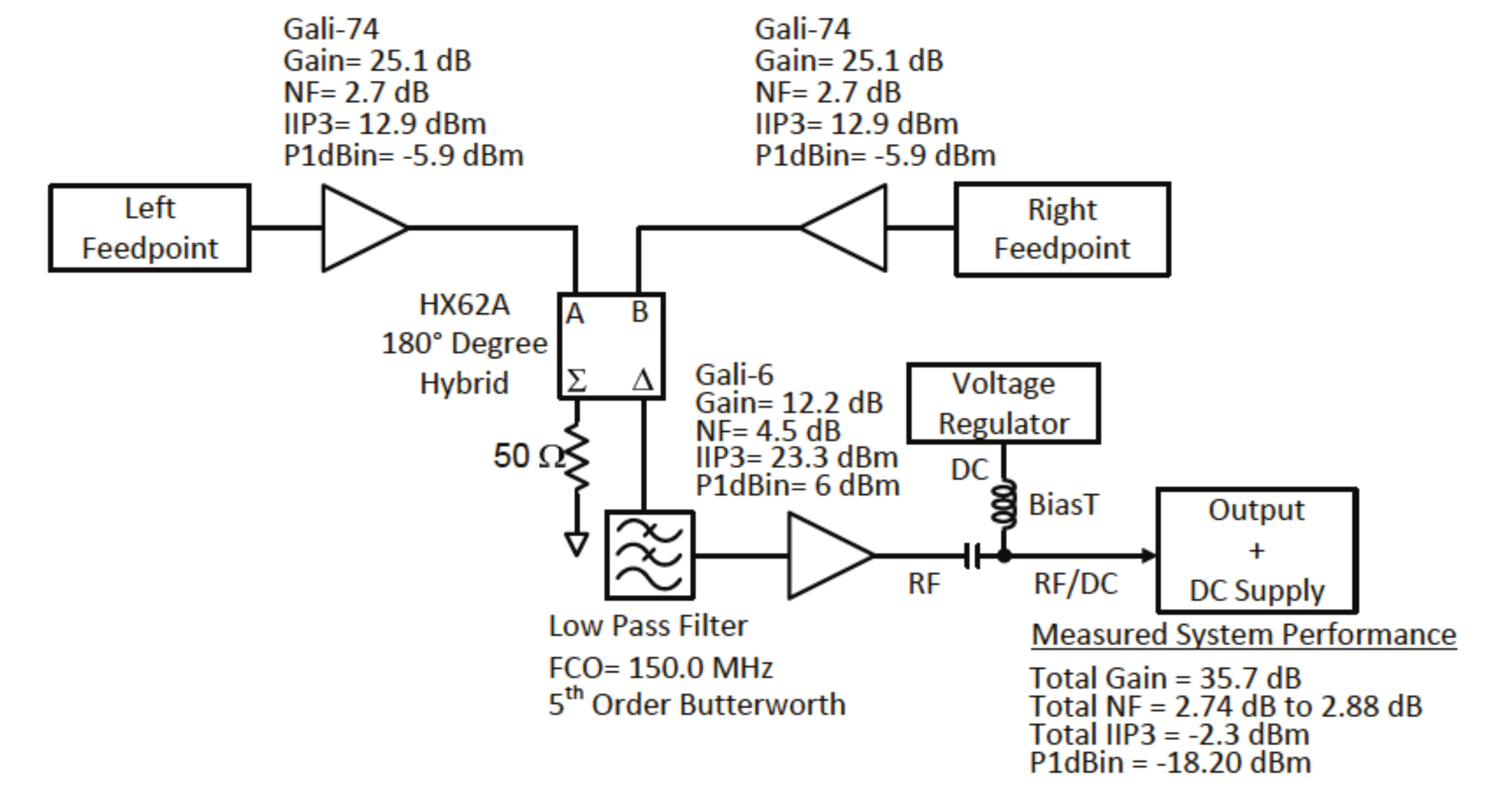}
\caption{Block diagram of one polarization of the LWA FEE electronics. The circuit schematic is shown in Figure \ref{FEE-Figure2-12}.}
\label{FEE-block}
\end{figure}

\begin{figure}
\plotone{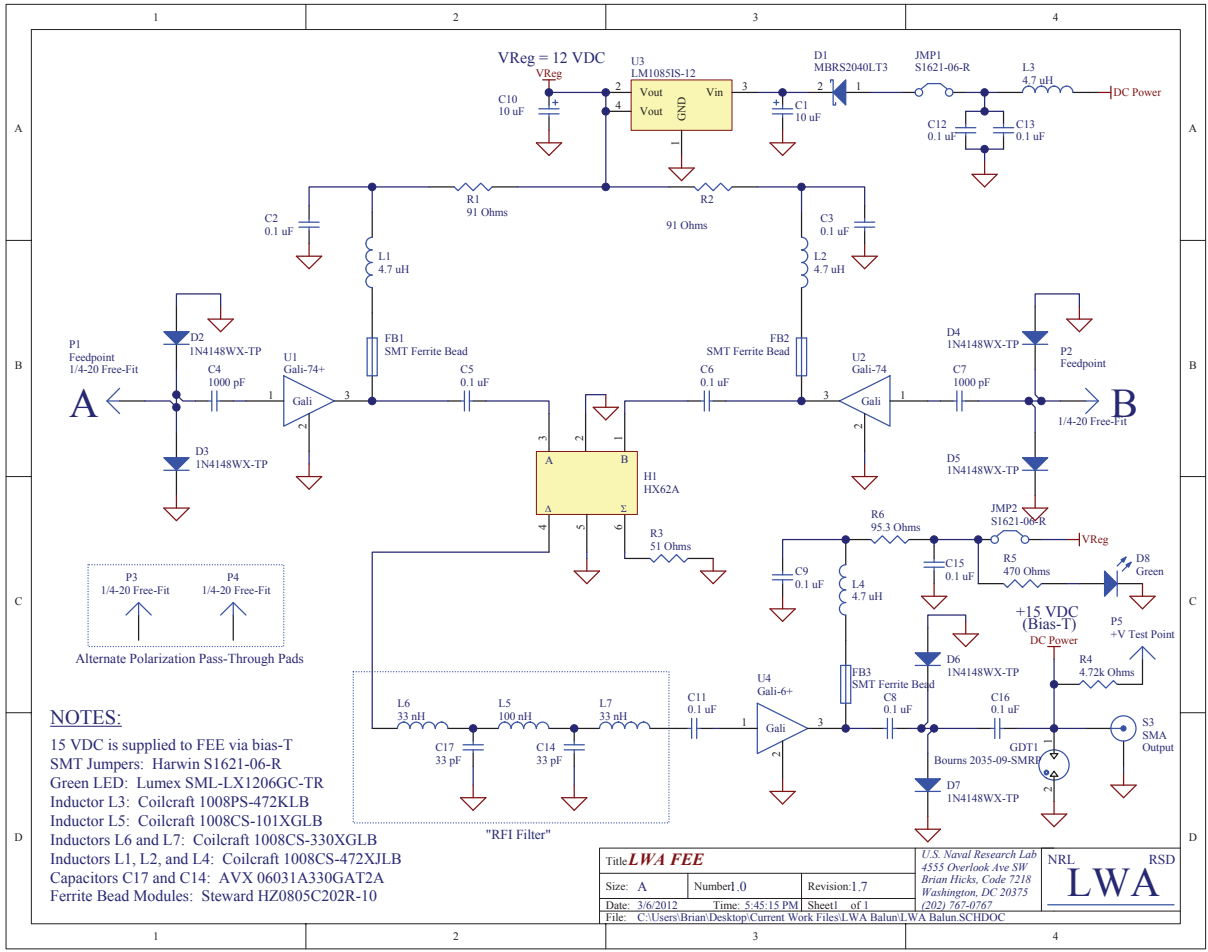}
\caption{Schematic of LWA Front End Electronics (FEE; Revision 1.7).}
\label{FEE-Figure2-12}
\end{figure}

The LWA FEE is an extension of the prototype active-balun design utilized for the Long Wavelength Demonstrator Array \citep[LWDA;][]{Bradley05,Lazio10}. The LWDA design already used low-cost Monolithic Microwave Integrated Circuits (MMICs), but improvements for the FEE described here include the use of an additional 12 dB of gain to handle cable losses without affecting noise performance  \citep{Hicks07}, a local voltage regulator, an integral $\rm 5^{\rm th}$ order Butterworth filter, transient protection (\eg, lightning protection), and direct feed point connections. The block diagram of the LWA FEE is shown in Figure \ref{FEE-block} and the circuit diagram of the final unit is shown in Figure \ref{FEE-Figure2-12}.

Dual polarization FEE units are formed by rotating two identical double-sided FEE circuit boards 90\degr ~and bolting them together back-to-back with ground planes touching. This geometry was motivated by the need for isolation between polarizations, serviceability, and economy of fabrication. 

\subsubsection{Input impedance\label{InputImped}}

The input impedance of the active balun ($Z_0$) is an important design parameter of the FEE, as it affects the bandwidth of the antenna system, the efficiency with which power is coupled into the antenna (see Figure \ref{ANT-Figure2-6}), and the mutual coupling with nearby antennas. Extensive studies based on optimized models and field measurements were undertaken to determine the optimal balun impedance \citep{Erickson03,Gaussiran05,Paravastu07a}.  High impedance baluns were initially considered because of their ability to buffer the widely varying dipole impedances over our relatively wide bandwidth. However, it was determined that raising the input impedance above 1 k$\Omega$ resulted in insufficient current flow into the balun,  making it impossible to maintain sky noise dominated operation \citep{Erickson03}.  Based on this early work, we began to optimize antenna topologies for desired beam pattern and a feedpoint impedance of approximately 100 $\Omega$ \citep{Paravastu07a}.  It is possible to obtain a feedpoint impedance of 100 $\Omega$ by directly buffering the individual feedpoint connections with inexpensive commercially available MMIC amplifiers exhibiting high input return loss.  A 180$\degr$ hybrid or transformer is then used to convert the amplifier outputs to a single ended 50 $\Omega$ output.  This method avoids the loss, and subsequent increase in noise temperature, associated with adding transformers and other matching networks before the first amplification stage, while lowering production costs.

\subsubsection{Filter design\label{Filter}}

A $\rm 5^{\rm th}$ order, low-pass Butterworth filter was included before the final 12 dB gain stage to define the bandpass and reject out-of-band interference that could drive the FEE into non-linear operation. The characteristics of the filter can be widely varied within the topology of the filter through component selection. The 3 dB point of the filter is at 150 MHz; at 250 MHz it achieves $\sim21$ dB of attenuation. A high cut-off frequency was chosen to minimize distortion of the working bandpass of 20 -- 80 MHz. 

\subsubsection{Performance\label{FEEperform}}

The FEE serves to fix the system noise temperature, match the antenna impedance to the coax signal cables running to the distantly located receiver, provide adequate gain to overcome cable loss, and limit out-of-band RFI presented to the analog receiver module. The performance of a single polarization of the FEE is given in Table \ref{tbl-FEE2-3}. A crossed polarization unit will draw twice as much current as a single FEE board for a total of 460 mA at 15 V DC. Total power consumption for a 256 element, crossed dipole station  is then $\sim1.8$ kW.

Environmental testing of the final design of the FEE was carried out by \cite{Hartman09b} between -20 - +40 $\degr$C, a temperature range not likely to be exceeded for an extended period at the LWA site. The gain dependence on temperature varies between Ð0.0042 dB/$\degr$C and Ð0.0054 dB/$\degr$C, with the magnitude of the slope monotonically increasing with frequency between 20 MHz and 100 MHz. The
phase also has a weak dependence on temperature, with a slope of Ð0.011 degrees/$\degr$C and Ð0.014 degrees/$\degr$C.

\subsubsection{Noise figure\label{NF}}

We measured the noise figure of the final FEE using an Agilent N9030A signal analyzer using an Agilent 346B noise source. The noise figure ranged from 2.74 dB (255K) to 2.88 dB (273K) over the frequency range of 20 -- 80 MHz. The N9030A signal analyzer estimates an intrinsic measurement uncertainty of 0.21 dB ($\sim15$ K).  We also measured the gain linearity and intermodulation distortion using one and two injected tones (see Hartman and Hicks 2009 for measurement details). The results of these measurements are presented in Table 3 and they agree closely with those predicted using an analytic cascade analysis\footnote{http://sourceforge.net/projects/rfcascade/} based on the data sheet values for the components. 

\subsubsection{Manufacturing\label{FEEmanufact}}

{\bf Manufacturing quotes:} We obtained manufacturing quotes from a number of companies interested in producing turn-key FEEs. While there was some variation, most quotes, including printed circuit board (PCB) fabrication, assembly, and the administrative overhead associated with ordering all of the requisite parts were $\sim \$200$ per polarization in 2009. 

{\bf Quality control and functional testing:} Test scripts to confirm basic functionality and conduct full characterization of an FEE are detailed in \cite{Hicks08}. We have discussed the basic functional test (gain, stability, power consumption) described in that document with manufacturers and they agree that it could be readily implemented as an automated test procedure. The FEE also includes a test point to allow proper supply voltage to be safely verified in the field after the FEE has been installed on the support stand.

\subsubsection{PCB layout and mechanical details\label{FEElayout}}

\begin{figure}
\plotone{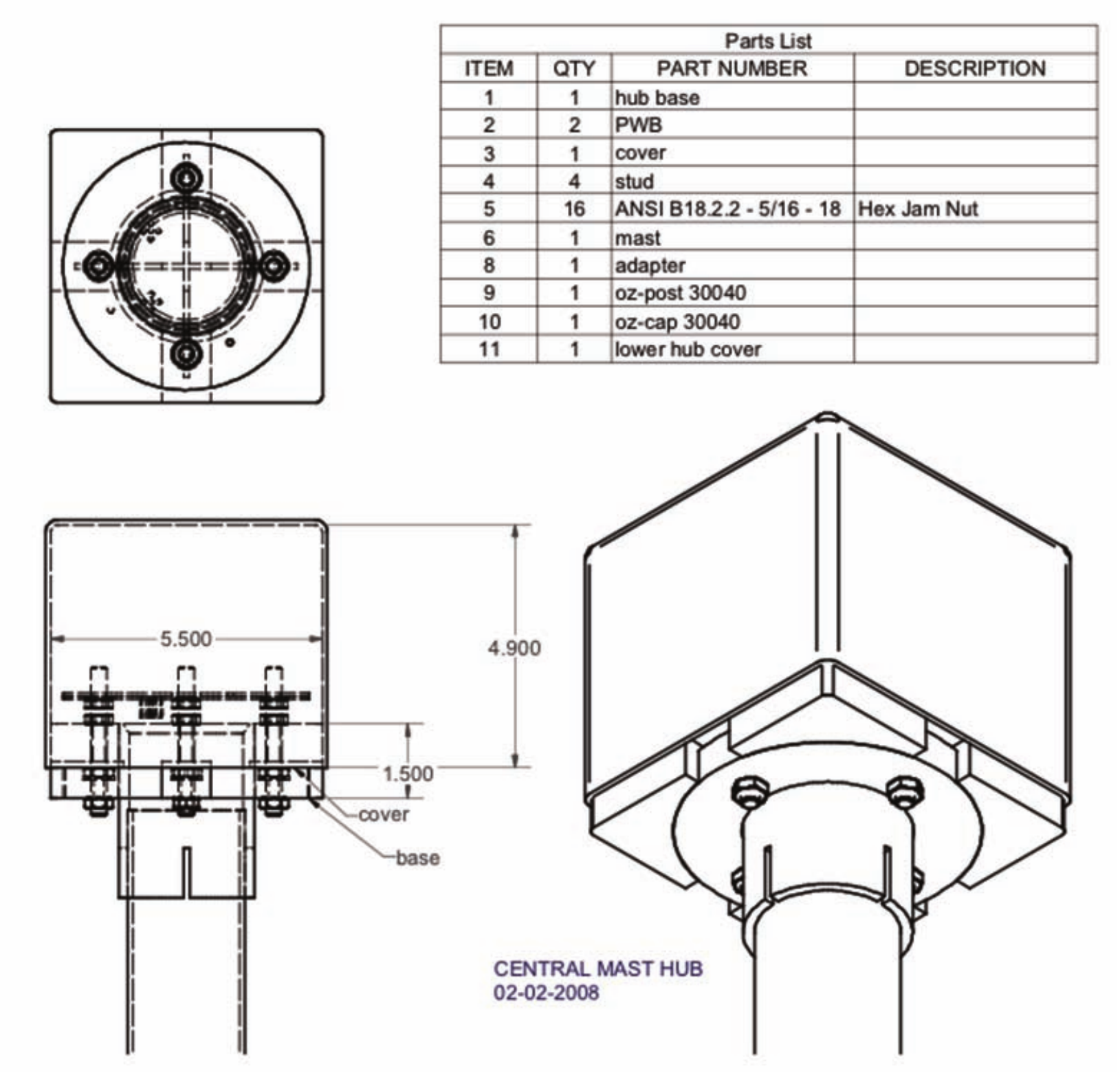}
\caption{Mechanical interface to the STD -- dimensions in inches.}
\label{FEE-Figure2-15}
\end{figure}

The components are all mounted on one side of the circuit board. The opposite side of the board is a solid copper ground plane aperiodically ``stitched'' to the grounded copper on the component side. The Bill of Materials for the FEE is given in  \cite{Hicks09}. The hard gold plated bolt holes on the FEE PCB directly connect to the stainless steel tabs connecting to the ANT dipole elements. Materials were chosen to avoid galvanic corrosion. The bolt holes were sized for 1/4-20 studs with standard clearance. A related mechanical interface to the STD was developed by Burns Industries of Nashua, NH\footnote{http://www.burnsindustriesinc.com} and is shown in Figure \ref{FEE-Figure2-15}. 
 
\subsubsection{Installation\label{FEEinstal}}

The FEE is installed on to the STD after the cables have been pulled and are ready to be connected. A keying scheme is incorporated into the FEE and STD hub such that the FEE can only be installed with the N/S polarization in the correct orientation. The connections to the coax cable are color coded for the two polarizations and 7 -- 10 in-lbs of torque is required to tighten the SMA connectors. 

 \subsection {Antenna stand - STD\label{STD}}

After considering several possible designs for the STD, we chose a central mast design. That conferred several advantages:

\begin{itemize}
\item The antenna elements are not required to be load bearing structural elements.
\item The antenna elements can be much easier to assemble than self-supporting pyramidal designs.
\item Site preparation work is minimized because the STD only touches the ground at one point.
\item The footprint of the design is smaller than the self-supporting pyramidal designs so there is more clearance between antenna systems.
\end{itemize}

\begin{figure}
\plotone{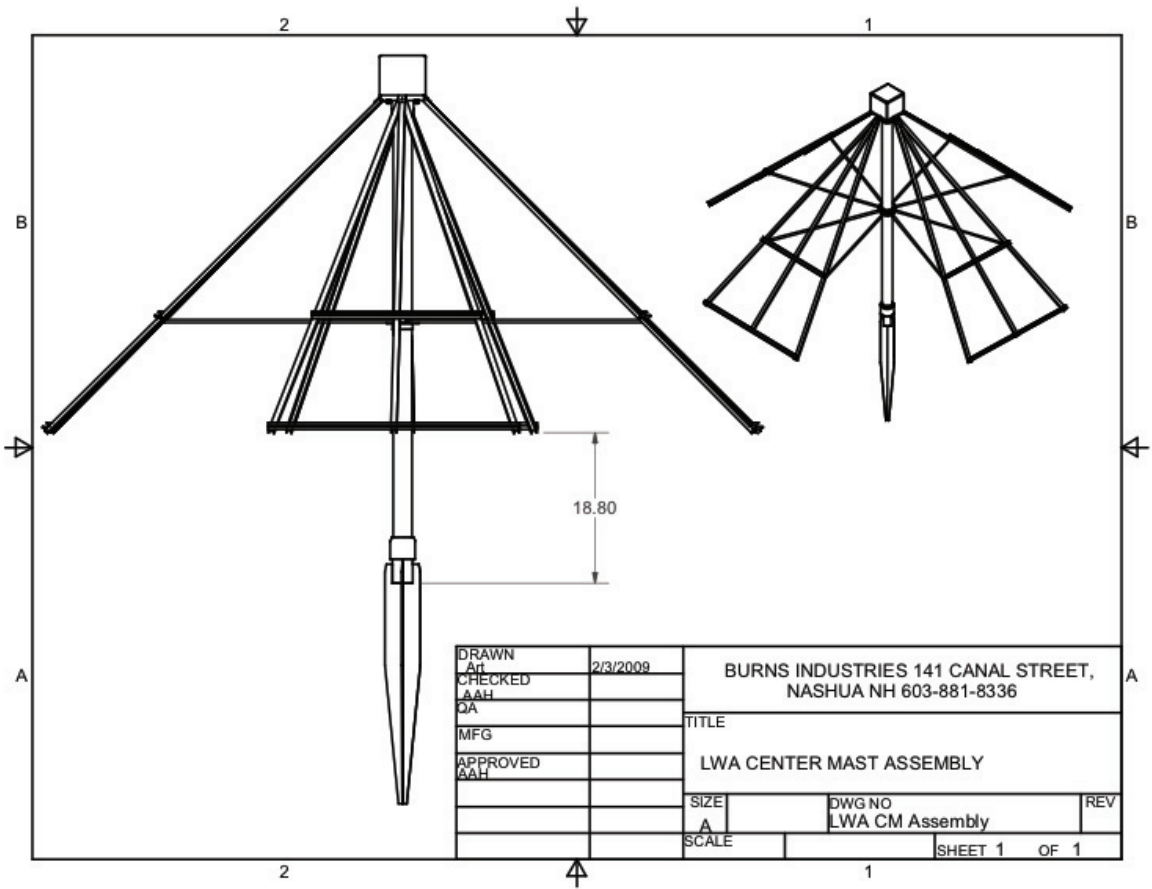}
\caption{Mechanical drawing of the STD structure. Dimensions are in inches.}
\label{STD-Figure2-8}
\end{figure}

We developed the central mast design in collaboration with our manufacturing partner Burns Industries, Inc.. The design is shown in Figure \ref{STD-Figure2-8}. It consists of four welded ``tied fork with crosspiece'' (see Section \ref{antgeom}) aluminum dipole arms attached to the bottom of a solid plastic hub at the top of the mast. The FEE is mounted to the top of the hub and the solid hub prevents mechanical stresses on the dipole arms from being transmitted through the stainless steel tabs to the FEE PCB. A plastic cap fits over the hub to protect the FEE from the elements. A fiberglass rod ``spider'' midway down the mast supports the dipole arms so that they do not move significantly in the wind. The mast is standard 2 3/8 inch ($\sim6$ cm) outer diameter galvanized steel fence post, machined to accept a mount to the junction box where the connection to the RF/Power Distribution (RPD) conduit is made. 

\subsubsection{Installation and alignment\label{STDinst}}

The components of the STDs were manufactured by Burns Industries, Inc. and shipped to the array site. The pieces were then assembled under a shelter and carried out to the mounting points. They were fitted with a compression collar and set into the Oz-Post\textregistered  \footnote{http://www.ozcobuildingproducts.com/Oz-Post.html} sleeve. After alignment (described below), the collar was hammered into place with the Oz-Post\textregistered ~CDT-07 - Cap Driving Tool\textregistered ~and the installation was complete. 

\begin{figure}
\plotone{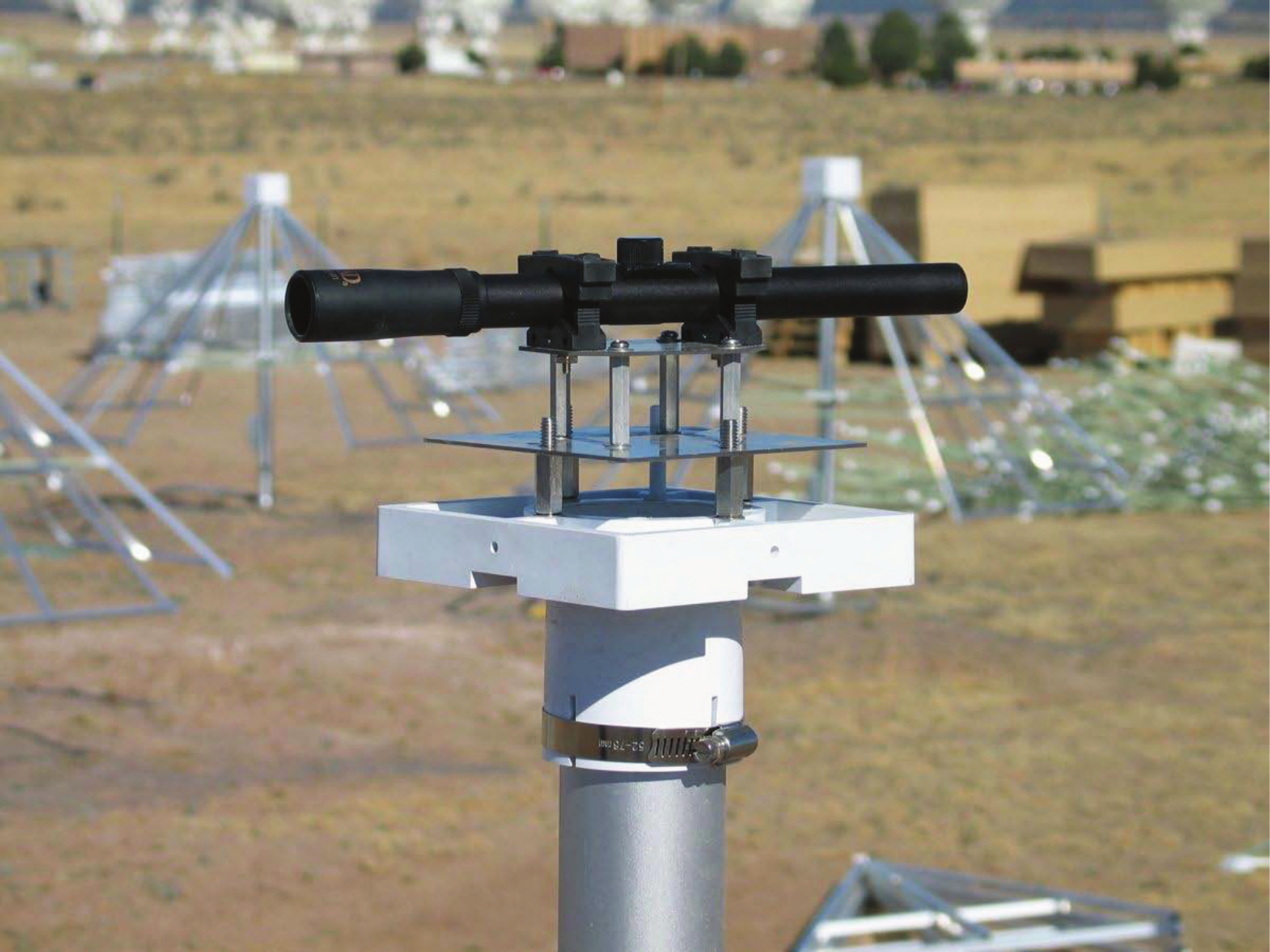}
\caption{STD sighting telescope alignment mechanism used for the LWA1.}
\label{STD-telescope}
\end{figure}

After field testing and finding that compass alignment of the STD with True North was unsatisfactory (possibly due to intrinsic magnetization in the steel mast), angular alignment of the LWA STDs was accomplished using a sighting telescope permanently fixed to a base that is identical in shape, polarization keying, and mounting holes to an FEE unit. The base provided a stable mount for an inexpensive 4X telescope commonly sold for use with air rifles (see Figure \ref{STD-telescope}).

Through surveying, the angular offset at the STD installation point from True North to a distant, geographic reference point was established in advance and the telescope was firmly mounted to the base with that offset. The antenna hub was then rotated until the distant reference appeared in the crosshairs of the sighting telescope, indicating that the hub was properly aligned and ready to be locked in place.  For the LWA1, the $\sim 40$ km distant peak of South Baldy Mountain served as the reference point and its azimuth was offset by $102\degr$ from True North. Clearly, the fixture must be site specific, but once manufactured it allowed rapid and precise alignment by untrained personnel.  Experience has shown that the procedure can easily produce alignment with True North to a tolerance of $<5\degr$, which is more stringent than the system requirement \citep[see][]{Janes09}. 

\subsubsection{Mechanical and environmental survivability\label{STDmech}}

The survivability requirements in ``The Long Wavelength Array System Technical Requirements'' document \citep{Clarke09,Janes09} include survival of winds up to 80 mph with gusts to 100 mph (wind speed up to 36 m s$^{-1}$ with gusts up to 45 m s$^{-1}$), UV lifetimes of 15 years, and alighting of a 4 lb ($\sim 2$ kg) bird. Both the fiberglass and plastic in the STD design are UV stabilized materials with long lifetimes. Wind survivability has been verified by both modeling and, so far, three years of testing in the field. No problem with heavy birds has been noted.

\subsubsection{Removal\label{STDremove}}

Using Oz-Posts\textregistered \  as the ground anchors facilitates removal of the STDs, should we need to return the site to its original condition. The Oz-Post\textregistered \ collars are removable, so the masts may be removed, and the Oz-Post\textregistered \ company sells a simple device, the ``Oz Post Oz Puller with Post Clamp\textregistered'' for pulling the posts out of the ground.

\subsection{Ground screen - GND\label{GND}}

\cite{Paravastu07b,Stewart07,York07} carried out simulations and detailed tests of the effects of deploying large and small ground screens above both wet and dry ground. They concluded that that there are significant benefits from deploying a ground screen beneath the antennas, including reduced ground losses and reduced susceptibility to variable soil conditions. Additionally, for an antenna in isolation, \cite{Kerkhoff07b} demonstrated that a small ground screen provides these benefits without the axial asymmetry and significant sensitivity to RFI coming from the horizon that are caused by using a full-station ground screen \citep{Paravastu07a,Schmitt08}. It is difficult to accurately model these effects for a full array in the presence of mutual coupling, but initial studies \citep{Kerkhoff07b,Ellingson09b} indicate that the behavior of a random array of antennas should be qualitatively similar to that of an antenna in isolation \citep{York07}.

\subsubsection{Design\label{GNDdesign}}

For the above reasons, we chose a 10 $\times$ 10 ft ($\sim3~\rm{x}~3$ m) ground screen under each STD, as detailed in \cite{Schmitt08} and \cite{Robbins09}. Simulations \citep{Robbins09} indicated  that the mesh density was not important as long as the lattice spacing was less than 12 inches ($\sim30$ cm). We chose a 4 $\times$ 4 inch ($\sim10~\rm{x}~10$ cm), galvanized welded wire mesh material that is structurally sound and inexpensive, made with wire diameter of 14 gauge ($\sim 2$ mm). A vendor was found that produces rolls of this material with dimensions of 6 $\times$ 200 ft ($\sim2~\rm{x}~60$ m). Considering that we needed two 6 $\times$ 10 ft ($\sim2~\rm{x}~3$ m) sections of mesh, overlapped by 2 ft ($\sim60$ cm), to make a 10 $\times$ 10 ft ($\sim3~\rm{x}~3$ m) ground screen, one of these rolls could be used to produce 10 complete ground screens. Taking into account possible mistakes and losses that can happen while cutting the mesh, we estimated a need for 27 rolls in order to produce 256 ground screens. 

For the physical connection of the two ground screen sections, we used split splicing sleeves (Nicopress\textregistered \  stock number FS-2-3 FS-3-4), 6 sleeves per ground screen (1,700 for a full 256 antenna station, assuming a 10\% loss). Simulations have shown \citep{Stewart09} that the performance of such a two-part ground screen is negligibly different from a single, unitary ground screen. The anchoring of the ground screens is also an important issue, since this must prevent the buckling of the sides of the mesh. For this purpose we used 12 inch ($\sim30$ cm) plastic tent stakes, 8 per ground screen, which were purchased in buckets containing 180 stakes each (12 buckets needed). 

\subsubsection{Installation\label{GNDinstal}}

The installation procedure of the ground screen was to unroll the mesh on a flat surface; cut it into 10 ft ($\sim3$ m) sections and flip each section upside down to prevent it from rolling back; overlap two 10 ft ($\sim3$ m) sections of mesh by 2 ft ($\sim60$ cm) and connect them using 6 splicing sleeves, spaced by 2 ft ($\sim60$ cm); and move the ground screens to the position of each stand and stake them, aligning the sides in the E/W by N/S direction with the ground screen centered on the Oz-Post\textregistered \ position. We then staked each corner of the ground screen and also put one stake in the mid point of each side to improve the stability.



\subsection {Antenna system performance\label{SystemPerform}}

\begin{figure}
\plotone{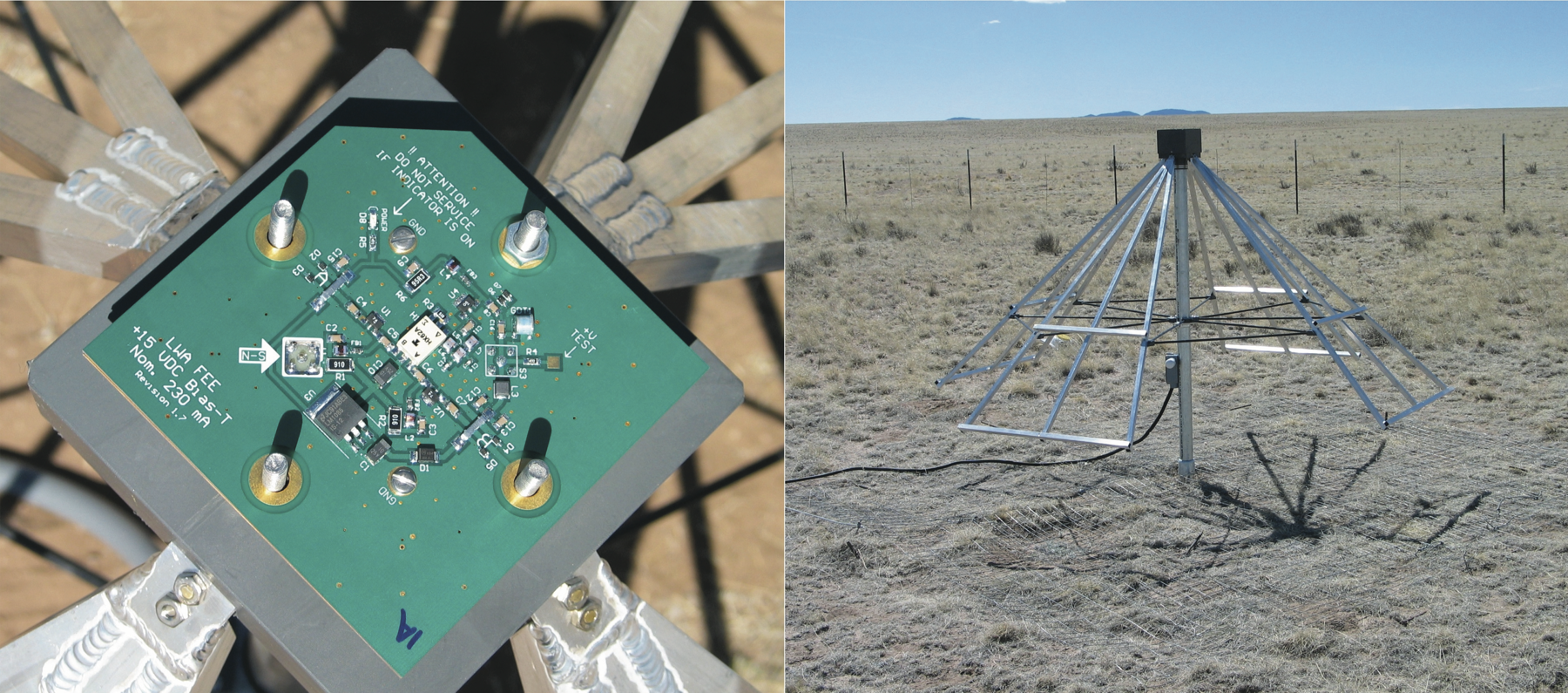}
\caption{(Left) Front End Electronics (FEE), without enclosure. (Right) FEE mounted at the center of the crossed ``tied fork with crosspiece'' dipole antenna (ANT), supported by the stand (STD), over the ground screen (GND) during a field test in April 2009 near the center of the NRAO's VLA site in western New Mexico. It should be noted that, even though this field test can be seen to be using cabling above ground, in order to prevent cable damage by animals, the final installation in LWA1 prohibited any exposed cables. All cables leave the antenna through electrical boxes to flexible conduit that mates to underground conduit. The cables then aggregate in closed junction boxes and the resulting bundles are brought back to the electronics in conduit.}
\label{FEE-Figure2-16}
\end{figure}

Of course, the ultimate question is how the antenna system performs in the field. In order to determine this, we performed initial tests with a preliminary prototype of the antenna system in July/August 2007 \citep{Paravastu08c}, carried out further tests with an improved prototype in September 2008, and then tested a final prototype of the ANT/FEE/STD/GND antenna system at our site in New Mexico on NRAO property near the center of the VLA in April 2009. A photo of the FEE (with the FEE protecting box removed) mounted on the STD with the crossed dipole ANT arms attached is shown on the left in Figure \ref{FEE-Figure2-16}. Shown in the same figure on the right is the assembled ANT/FEE/STD with the ground screen (GND) installed. \cite{Hartman09a} describes field measurements in April 2009 of two of the prototype antennas operating as an interferometer. 

\begin{figure}
\plotone{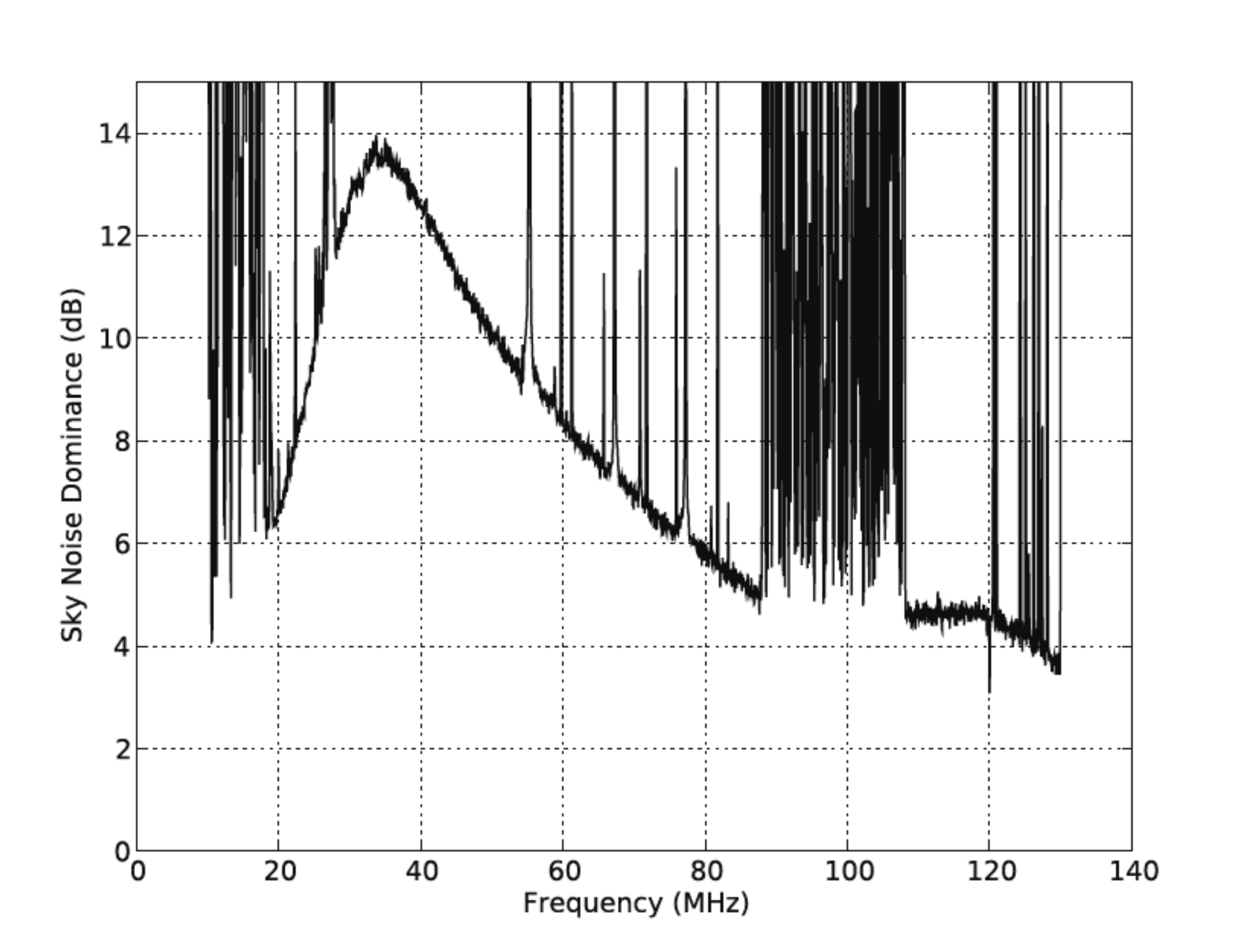}
\caption{Measured sky noise dominance (SND)  as a function of frequency for the actual antenna system described here. The system includes the  antenna (ANT; see Section \ref{ANT}), Front End Electronics (FEE; see Section \ref{FEE}), antenna stand (STD; see Section \ref{STD}) and ground screen (GND; see Section \ref{GND}). It should be noted that the SND meets or exceeds the requirement  that the system noise be dominated by the Galactic Background (sky noise) by at least 6 dB from 20 -- 80 MHz. Note also that, because the measurement was carried out on the sky, radio frequency interference (RFI) is visible.}
\label{ANT-Figure2-7}
\end{figure}

Most exciting, of course, is that the results of the tests of the full antenna system on the sky [see Figure \ref{ANT-Figure2-7}, \cite{Taylor12}, and \cite{Ellingson12}] show that it meets its requirement of $>6$ dB Sky Noise Dominance (SND) across the 20 -- 80 MHz band [Specification TR-10A listed in the technical requirements documents  \cite[see, \eg,][]{Kassim05b,Clarke09,Janes09}] and clearly shows sensitivity to the Galactic background emission (and interference) . 

Thus, the ANT/FEE/STD/GND antenna system meets requirements. This led directly to the installation of 256 examples for the LWA1 station described in Section \ref{LWA}.

\section{Summary and conclusions}

We have presented the mechanical and electrical design and response of a HF/VHF antenna system suitable for economical use in large numbers and currently deployed 256-fold in the Long Wavelength Array first station (LWA1) near the center of NRAO's VLA. The antenna system consists of crossed-dipole antennas (ANT), an active balun/amplifier front-end electronics (FEE), a supporting stand (STD), and a ground screen (GND). This system permits Galactic Background limited observations by at least 6 dB over the band from 20 -- 80 MHz with full sky coverage.

Because it must be reproduced $>10,000$-fold for a state-of-the-art long wavelength synthesis array in an unprotected desert environment, we have designed and procured an antenna system that is inexpensive, robust, easily deployed, and easily maintained with excellent response properties for radio astronomy synthesis imaging.  



\acknowledgments

We are very grateful to the National Radio Astronomy Observatory for their long-term support for long wavelength radio astronomy. Basic research in radio astronomy at the Naval Research Laboratory is supported
by 6.1 base funds.

\clearpage

\clearpage

\clearpage

\clearpage

\clearpage

\clearpage

\clearpage

\clearpage

\clearpage

\clearpage

\clearpage

\clearpage

\clearpage

\begin{table}
\begin{center}
\caption{LWA Technical Specifications  \citep[taken from][]{Kassim05b}\label{tbl-LWA-spec}}
\begin{tabular}{ll}
\tableline\tableline
Parameter & Design Goal\\
\tableline
Frequency range (minimum) & 20 -- 80 MHz\\
Effective collecting area & $\sim 1$ km$^2$ at 20 MHz\\ 
Number of dipoles & $\sim13,000$\\
Number of stations & $\sim50$\\
Station diameter & 100 m E/W $\times$ 110 m N/S\\
Crossed dipoles stands per station & 256 \\
Configuration & Core: 17 stations in 5 km\\
                          & Intermed.: 17 sta. in 5-50 km\\ 
                          & Outliers: 18 sta. in 50-400 km\\ 
Baselines & 0.2 -- 400 km\\
Point-source sensitivity & $\sim1.1$ mJy at 30 MHz\\
~~~~~(2 pol., 1 hr integ., 4 MHz BW) & $\sim0.7$ mJy at 74 MHz\\
Sky Noise Dominance (SND) & $\ge6$ dB from 20 -- 80 MHz \\
Maximum angular resolution & $\sim5''$ at 30 MHz\\
                         & $\sim2''$ at 74 MHz\\ 
Station Field of View (FoV) & $\sim2$\degr at 74 MHz \\
Number of independent FoV & 2 -- 8\\
Mapping capability & Full FoV\\
Maximum observable bandwidth & 32 MHz\\
Spectral resolution & $<1$ kHz\\
Time resolution & 1 ms\\
Image dynamic range & $>10,000$\\
Polarization & Full Stokes\\
Digitized bandwidth & Full RF\\
\tableline
\end{tabular}
\end{center}
\end{table}
 
\clearpage

\begin{table}
\begin{center}
\caption{Simulated antenna power pattern summary. Values are zenith angles at which the power pattern is down by 3 dB and 6 dB from the zenith gain. \citep[taken from][]{Hicks09}\label{tbl-ANTpatt}}
\begin{tabular}{cccccc}
\tableline\tableline
Frequency & Gain & \multicolumn{2}{c}{E-plane} & \multicolumn{2}{c}{H-plane} \\
                   &            & \multicolumn{2}{c}{------------------} & \multicolumn{2}{c}{------------------} \\
                   & (dBi) &  -3 dB & -6 dB & -3 dB& -6 dB\\
\tableline
20 MHz & 4.0 & 41\degr & 57\degr & 51\degr & 66\degr \\
40 MHz & 6.0 & 45\degr & 64\degr & 53\degr & 67\degr \\ 
60 MHz & 5.9 & 48\degr & 71\degr & 55\degr & 68\degr \\
80 MHz & 5.6 & 45\degr & 77\degr & 58\degr & 70\degr \\
\tableline
\end{tabular}
\end{center}
\end{table}

\clearpage

\begin{table}
\begin{center}
\caption{Measured FEE performance summary\label{tbl-FEE2-3}}
\begin{tabular}{lc}
\tableline\tableline
Parameter & Value\\
\tableline
Current Draw (at +15 VDC) & 230 mA \\
Voltage Range & $\pm 5$\% \\
Gain  & 35.5 dB \\
Noise Temperature & 255 - 273 K  \\ 
Input 1 dB Compression Point &  $-$18.20 dBm  \\
Input $3^{\rm rd}$ order intercept (IIP3) & $-$2.3 dBm \\
\tableline
\end{tabular}
\end{center}
\end{table}



\end{document}